\documentclass[%
 reprint,
 amsmath,amssymb,
 aps,
floatfix,
]{revtex4-1}

\usepackage{graphicx}
\usepackage{dcolumn}
\usepackage{bm}
\usepackage{color}
\usepackage{float}
\begin{document}


\title{Molecular dynamics investigation of soliton propagation in a\\ 
two-dimensional Yukawa liquid}

\author{Z. Donk\'o}
\affiliation{Institute for Solid State Physics and Optics,
Wigner Research Centre for Physics,\\
H-1121 Budapest, Konkoly-Thege Mikl\'os str. 29-33, Hungary}

\author{P. Hartmann}
\affiliation{Institute for Solid State Physics and Optics,
Wigner Research Centre for Physics,\\
H-1121 Budapest, Konkoly-Thege Mikl\'os str. 29-33, Hungary}

\author{R. U. Masheyeva}
\affiliation{IETP, Al-Farabi Kazakh National University, 71 Al-Farabi av., Almaty 050040, Kazakhstan}

\author{K. N. Dzhumagulova}
\affiliation{IETP, Al-Farabi Kazakh National University, 71 Al-Farabi av., Almaty 050040, Kazakhstan}

\date{\today}

\begin{abstract}
We investigate via Molecular Dynamics simulations the propagation of solitons in a two-dimensional many-body system characterized by Yukawa interaction potential. The solitons are created in an equilibrated system by the application of electric field pulses. Such pulses generate pairs of solitons, which are characterized by a positive and negative density peak, respectively, and which propagate into opposite directions. At small perturbation, the features propagate with the longitudinal sound speed, form which an increasing deviation is found at higher density perturbations. An external magnetic field is found to block the propagation of the solitons, which can, however, be released upon the termination of the magnetic field and can propagate further into directions that depend on the time of trapping and the magnetic field strength.  
\end{abstract}

\maketitle

\section{\label{sec:intro}Introduction}

With the pioneering observation of ``wave of translation'' made by John S. Russell in 1834 on the Union Canal in Scotland, the field of non-linear wave phenomena was born. Russel continued his investigations in water channel experiments by triggering waves with translating vertical plates placed in the water \cite{Russel1838,Russel1845}. With his experiments he was able to determine some properties of the single (``solo'' or as later called ``solitary'') waves, like that those are stable features, which can travel a long distance, and that the speed of the wave depends on its amplitude and on the water depth. He also found that two waves do not disturb each other so that, e.g., they can overtake each other.

Due to the lack of proper theoretical description of non-linear wave phenomena the field did not experience much progress for several decades. This changed with the derivation of the Korteweg--de Vries (KdV) equation in 1896 \cite{Korteweg1896}, which is the simplest equation embodying non-linearity and dispersion. Despite its apparent simplicity it is very rich and has a large variety of solutions, including spatially localized solitary waves and periodic cnoidal waves. In the meantime, solitons have been identified in many different fields beyond hydrodynamics, including optics (optical fibers and non-liner media) \cite{Hasegawa73}, magnetism \cite{Kosevich1998}, nuclear physics \cite{Iwata19}, and Bose-Einstein condensates \cite{Frantzeskakis10}.

Solitons are salient non-linear features in plasmas too. A review of the early experimental findings of ion-acoustic solitons was published in \cite{Tran79}. The combination of large family of systems we call plasmas and the richness of the KdV equation and its variants provides seemingly unlimited possibilities for theoretical investigations. Just to mention some of the most recent ones, the collision properties of overtaking small-amplitude super-solitons, as well as solitons with opposite polarities in a plasma consisting of cold ions and electrons with two-temperature Boltzmann energy distributions were investigated in \cite{Olivier18,Verheest19}. The effect of relativistic corrections to the electron kinetics on wave propagation was discussed in \cite{EL-Shamy19}. Ion acoustic solitons in multi-ion plasmas have been analyzed in \cite{Ur-Rehman19,Alam19}. Multiple soliton solutions have been presented in (3+1) dimensions in \cite{Wazwaz15}. The effect of magnetic field acting on the dust particle motion was taken into account in \cite{Saini16,Atteya18,Yahia19}, while solitary waves and rogue waves in a plasma with non-thermal electrons featuring Tsallis distribution have been derived in \cite{Wang13}. Bending of solitons in weak and slowly varying inhomogeneous plasmas was shown in \cite{Mukherjee15} and the  application of solitary waves for particle acceleration has been discussed in \cite{Ishihara18}. 

More closely related to the topic of this paper, in the field of strongly coupled dusty plasma research solitons became of high interest in the last decade after the pioneering experiments of Samsonov et. al. \cite{Samsonov02}. Numerous single layer dusty plasma experiments have followed providing more data and insight \cite{Nosenko04,Nosenko06,Sheridan08}. The connection between wave amplitude, width and propagation velocity has been explored in \cite{Bandyopadhyay08}, the existence of dissipative dark (rarefactive) solitons has been reported in \cite{Heidemann09,Zhdanov10}. Experiments on the collision of solitons were presented in \cite{Boruah15}. Experimental observations on the modifications in the propagation characteristics of precursor solitons due to the different shapes and sizes of obstacles over which the dust fluid had to flow was presented in \cite{Arora19}. In three-dimensional dusty plasmas under micro-gravity conditions solitons could be launched as well, as reported in \cite{Usachev14}. Solitary waves in one-dimensional dust particle chains were studied in \cite{Sheridan17}.

Supporting and extending experimental possibilities concerning solitons in dusty plasma, theoretical studies focused on the effects of charge-varying dusty plasma with nonthermal electrons \cite{Berbri09}, of an external magnetic field \cite{Nouri12}, of the external periodic perturbations and damping \cite{Chatterjee18}, as well as of the presence of dust particles with opposite polarities \cite{Rahman18}. The effects of dust--ion collision \cite{Paul19}, and the possibility of cylindrical solitary waves \cite{Gao19} were also addressed. 

Numerical simulations became as well very useful for exploring the physics of strongly coupled dusty plasmas. The most widely used approach has been the molecular dynamics (MD) method, which solves the equations of motion of the particles with forces originating from the specified pairwise inter-particle potential and from any external forces, if present. In many settings, the validity of the Newtonian equations of motion can be assumed. In case of an appreciable interaction between the particles and the embedding gaseous system, the Langevin simulation approach \cite{Langevin} can be adopted. Recent molecular dynamics simulations have shown the presence of solitary waves and their compatibility with the predictions of the KdV equation \cite{Avinash03,Kumar17}, the possibility of excitation of solitons with moving charged objects \cite{Tiwari16}, and the presence of rarefactive solitons \cite{Tiwari15}.

Since the first laboratory plasma crystal experiments in 1994 \cite{Chu94,Thomas94,Melzer94} strongly coupled two-dimensional dusty plasmas and the corresponding single layer Yukawa one-component plasma model became popular model systems for the investigation of various structural, transport, and dynamical properties of many-body systems. These systems provide the unique possibility to observe collective and many-body phenomena on the microscopic level of individual particles. Most recent works on transport processes include studies of the effect of external magnetic fields on the different transport parameters \cite{Baalrud17,Feng17,Hartmann19} and testing different thermal conductivity models with equilibrium molecular dynamics simulation \cite{Scheiner19}. Related to waves and instabilities, the coupling of non-crossing wave modes has been addressed in \cite{Meyer17}, spiral waves in driven strongly coupled Yukawa systems were analysed in \cite{Kumar18}, the effect of periodic substrates were studied in \cite{Li18,Wong18}, thermoacoustic instabilities in the liquid phase were shown in \cite{Kryuchkov18}, and microscopic acoustic wave turbulence was analyzed in \cite{Hu19}. Studies on structural phase transitions included freezing \cite{Hartmann10,Su12} and melting \cite{Petrov2015,Jaiswal_2019}. 

Dusty plasmas also provide an easily accessible playground for testing new theoretical approaches and concepts. E.g., the shear modulus for the solid phase was obtained from the viscoelasticity in the liquid phase \cite{Wang19}, a survival-function analysis was performed to identify multiple timescales in strongly coupled dusty plasma \cite{Wong18}, the applicability of the configurational temperature in dusty plasmas was investigated \cite{Himpel19}. In ref. \cite{Choi19} high-precision molecular dynamics results were used for testing of theoretical models of ion-acoustic wave-dispersion relations and related quantities. Studies at the mesoscopic scale of finite dust particle clusters are probably at the most fundamental level, as the contribution of every individual particle is significant. Most recently amplitude instability, phase transitions, and dynamic properties were studied \cite{Lisina19}. In addition to fundamental many-body physics dusty plasmas provide a very sensitive tool for the detailed investigation of mutual plasma--surface interactions. Recent studies include the investigations of the effect of external fields on the local plasma properties around a dust particle \cite{Sukhinin17}, and the sputtering rate of the solid dust surface with nanometer resolution in \cite{Hartmann17}.

In this work, we present molecular dynamics investigations of the propagation of solitons in a 2-dimensional strongly-coupled many body system, characterized by Yukawa pair potential. The solitons are created by  electric field pulses. Their propagation and their collisions are traced at various system parameters. The effect of an external magnetic field is also addressed. In Sec. \ref{sec:model} we describe our simulation techniques, the generation and the characterization of the solitons. In Sec. \ref{sec:results} we report the results of our studies, while in Sec. \ref{sec:summary} a brief summary of our findings is given.

\section{Simulation technique}

\label{sec:model}

The system that we investigate consists of $N$ particles that have the same charge, $Q$, and mass, $m$, and reside within a square computational box, to which we apply periodic boundary conditions. The edge length of the box is $L$ that results a surface density of the particles $n = N / L^2$ and a Wigner-Seitz radius of $a = (\pi n)^{1/2}$. The particles interact via the screened Coulomb (Debye-H\"uckel, or Yukawa) potential 
\begin{equation}
\phi(r) = \frac{Q}{4 \pi \varepsilon_0}~\frac{\exp(-r /  \lambda_{\rm D})}{r},
\label{eq:pot}
\end{equation}
where $\lambda_{\rm D}$ is the screening (Debye) length, which depends on the characteristics of the plasma environment (densities and temperatures of the electrons and ions) that embeds the dust system. We account for the presence of the plasma environment solely by taking into account its screening effect, i.e., we assume that the friction force that originates from this plasma environment is negligible. Correspondingly, we use the Newtonian equation of motion to follow the trajectories of the particles ($i=1,2,\dots,N$), 
\begin{equation}
m \ddot{\bf r}_i = {\bf F}_i = Q\,{\bf v}_i \times {\bf B} + \sum_{i \neq j} {\bf F}_{ij}(r_{ij}) + Q\,{\bf E}^\ast,
\label{eq:eom}
\end{equation}
where ${\bf B}$ is an external magnetic field that is perpendicular to the simulation plane, ${\bf F}_{ij}(r_{ij})$ is the force that acts on particle $i$ due to its interaction with particle $j$ situated at a distance $r_{ij}$, while the last term represents the force originating from the electric field ${\bf E}^\ast$ that is used to create the density perturbations. Our studies cover both $\textbf{B}=0$ and $\textbf{B} \neq 0$ cases. The characteristics of ${\bf E}^\ast$ will be defined below.

The summation for the inter-particle forces is carried out within a domain limited by a cutoff distance, $r_{ij} \leq r_{\rm c}$. The exponential decay of the pair potential allows us to assume that force contributions from particles at $r_{ij} > r_{\rm c}$ are negligible. $r_{\rm c}$ is set in a way to ensure that $F(r_{\rm c}) / F(2a) \sim 10^{-5}$ (recall that the nearest neighbour distance is $\approx 2a$). Finding the particles that give a contribution to the force acting on a given particle is aided by the chaining mesh technique. The resolution of the chaining mesh is set to be equal to $r_{\rm c}$.

In the cases when no magnetic field is applied, the equations of motion are integrated using the Velocity-Verlet scheme with a time step $\Delta t$ that provides a good resolution and accuracy over the time scale of the inverse plasma frequency $\omega_{\rm p}^{-1} = (n Q^2 / 2 \varepsilon_0 m a)^{-1/2}$ (by setting $\omega_{\rm p} \Delta t = 1 /30$). In the presence of magnetic field, the integration of the equations of motion is based on the method described in \cite{SPREITER1999102}.

The system is characterised by three dimensionless parameters: the coupling parameter
\begin{equation}
\Gamma = \frac{Q^2}{4 \pi \varepsilon_0}~\frac{1}{a k_{\rm B} T},
\end{equation}
where $T$ is the temperature, the screening parameter
\begin{equation}
\kappa = a / \lambda_{\rm D},
\end{equation}
and the normalised magnetic field strength
\begin{equation}
\beta = \Omega_{\rm c} / \omega_{\rm p},
\end{equation}
where $\Omega_{\rm c} = QB/m$ is the cyclotron frequency.

Upon the initialisation of the simulations the particles are placed at random positions within the simulation box and their velocities are sampled from a Maxwell-Boltzmann distribution that corresponds to the temperature defined by $\Gamma$. During the first 20\,000 time steps the system is equilibrated by re-scaling in each time step the velocities of the particles to match the desired system temperature. As this type of 'thermalisation' produces a non-Maxwellian velocity distribution, no measurements on the system are taken during this initial phase of the simulation. This phase is followed by a 'free run' period (consisting of 10\,000 time steps), during which the system is no longer thermostated and is allowed to equilibrate due to the interaction between the particles. 

Following this phase, solitons in the system are created by applying an electric field pulse with a duration of $\omega_{\rm p} \Delta t = 1 /6$ and having a spatial form 
\begin{equation}
E^\ast(x) = \widetilde{E}_0 \, \frac{Q}{4 \pi \varepsilon_0 a^2} \exp \biggl[-\frac {(x - x_0)^2}{2 w^2} \biggr],
\label{eq:field}
\end{equation}
where $x_0$ is the position where the soliton is to be generated, and $w$ the width of the perturbation region. We set this value to $w = 0.01 L$. The pulse is spatially homogeneous in the $y$ direction, i.e. particles with a given $x$ coordinate experience the same force regardless of their $y$ coordinates.  $\widetilde{E}_0$ in eq.\,(\ref{eq:field}) is a dimensionless scaling factor that controls the strength of the perturbation:  the factor $Q / (4 \pi \varepsilon_0 a^2)$ ensures that at $\widetilde{E}_0 = 1$ the peak value of the perturbing force acting on a particle at $x_0$, $E^\ast Q$, equals the Coulomb force between two particles separated by a distance $r=a$. 

\begin{figure}[!h]
\begin{center}
\footnotesize(a)\includegraphics[width=0.4\textwidth]{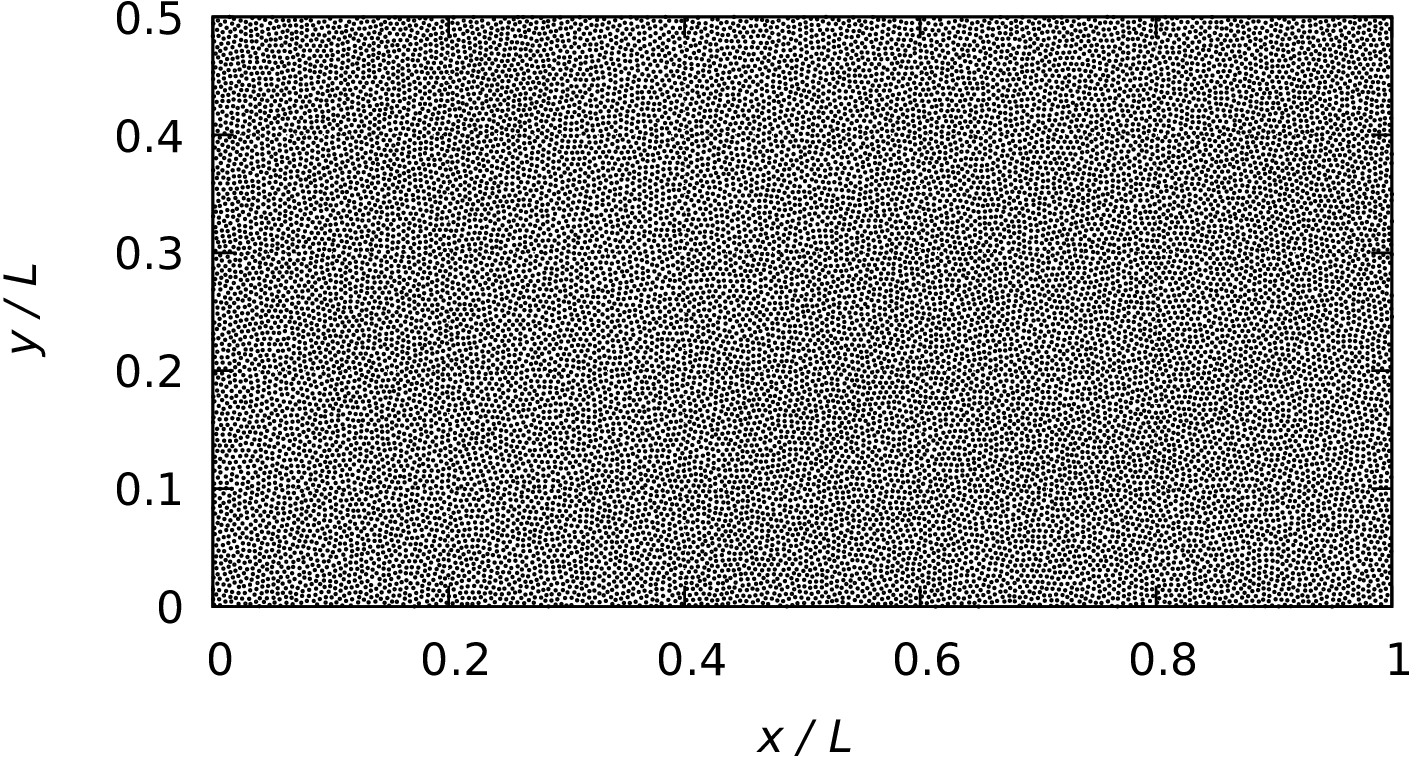}\\
\footnotesize(b)\includegraphics[width=0.4\textwidth]{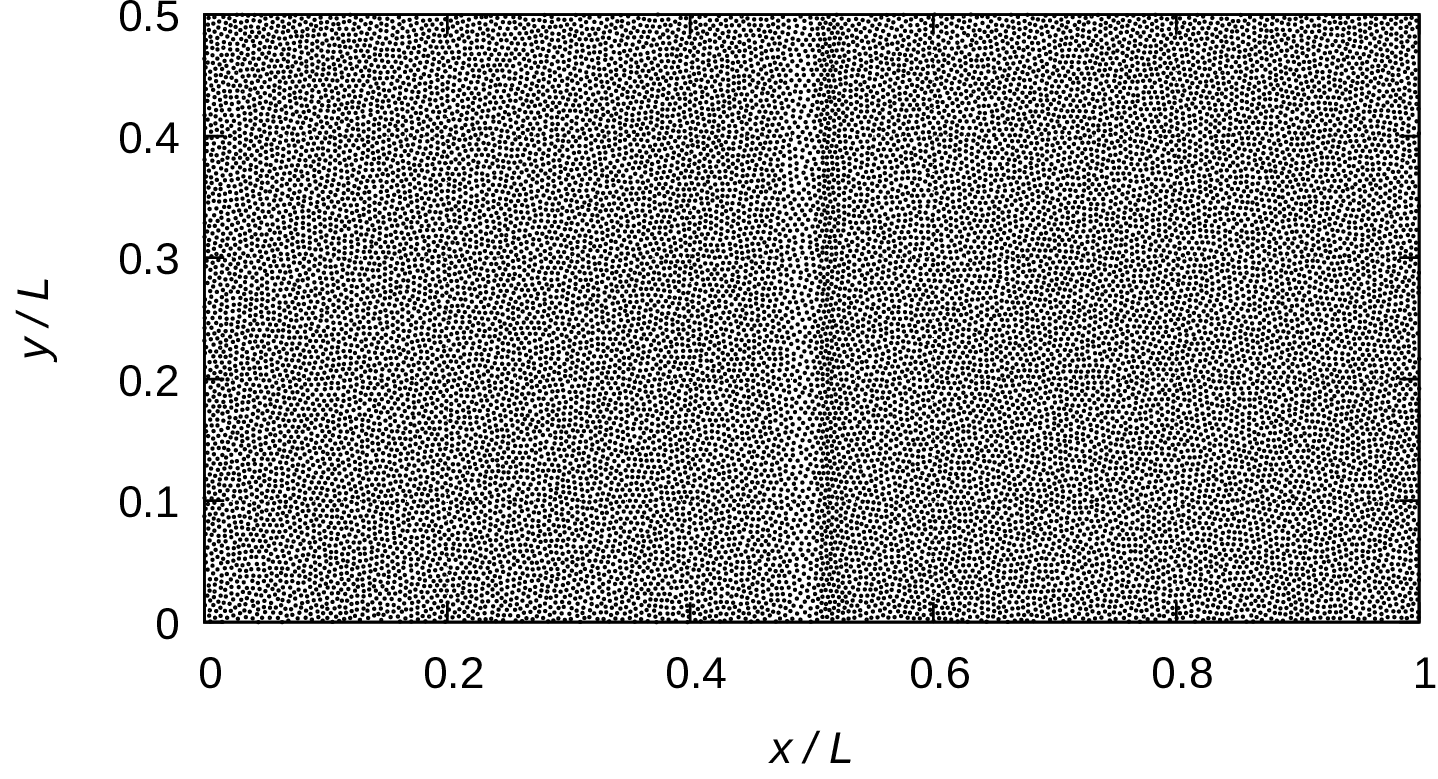}\\
\footnotesize(c)\includegraphics[width=0.4\textwidth]{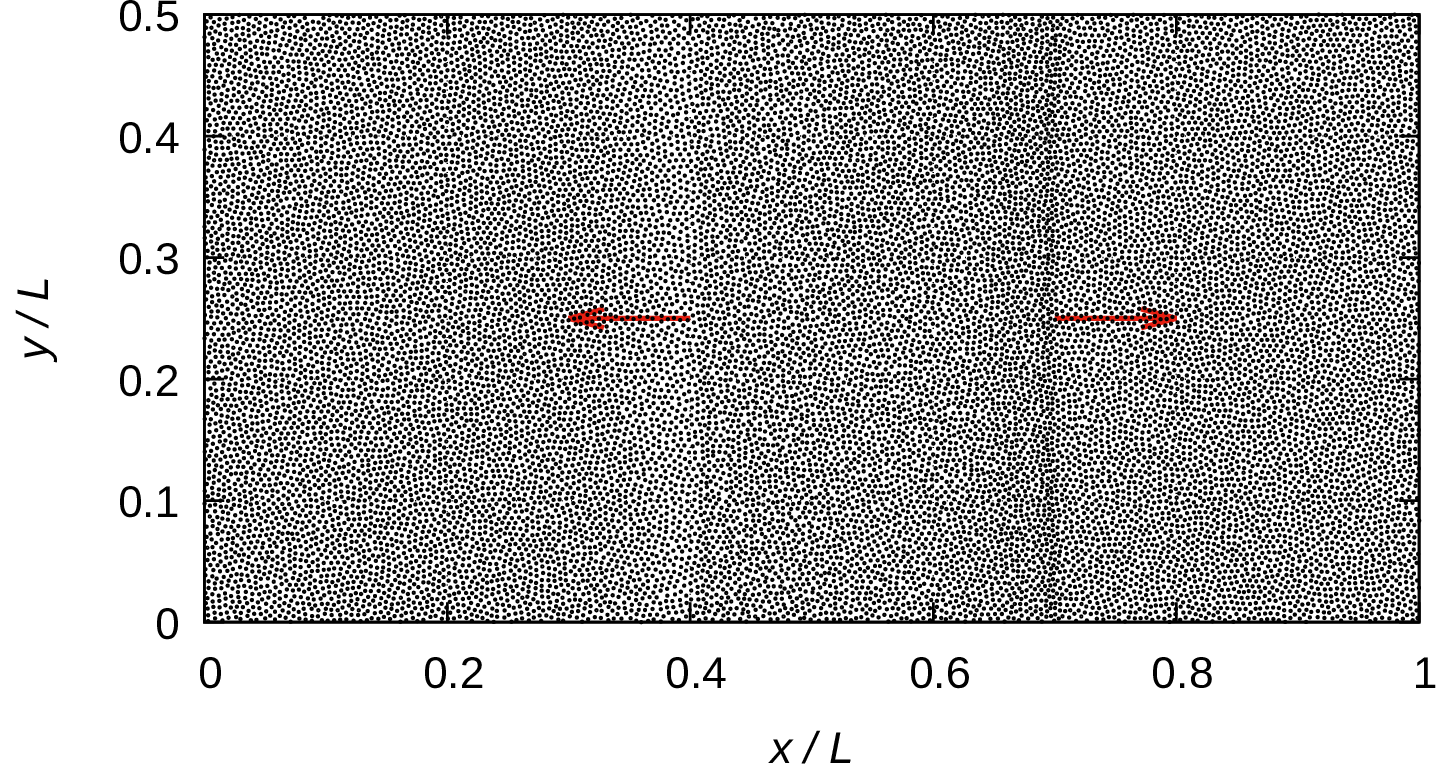}\\
\footnotesize(d)\includegraphics[width=0.4\textwidth]{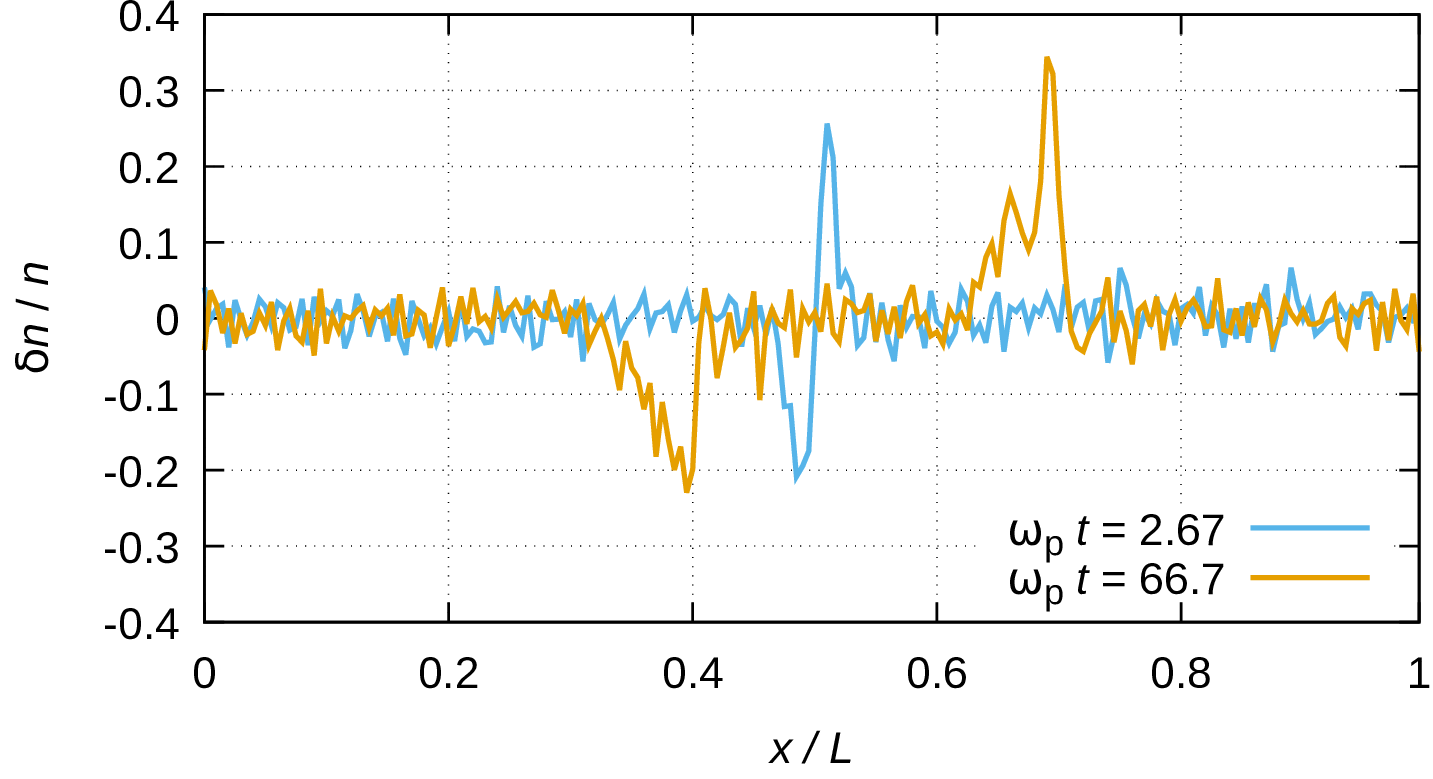}
\caption{Snapshots of the system at (a) $\omega_{\rm p} t$ = $-66.7$ (equilibrium phase, before perturbation), (b) $\omega_{\rm p} t$ = 2.67 (right after perturbation), and (c) $\omega_{\rm p} t$ = 66.7 (well after perturbation). The perturbation electric field is applied at the time $t$ = 0 and at the position  $x_0 = L/2$ with an amplitude $\widetilde{E}_0$ = 8.3. The red arrows in (c) indicate the direction of propagation of the density fronts. A positive density perturbation propagates in the $+x$ direction, while a negative density perturbation propagates in the $-x$ direction. (d) The density of the particles (averaged over the $y$ direction) at $\omega_{\rm p} t$ = 2.67 and 66.7. The plots in (a-c) show only half of the system in the $y$ direction. $\Gamma=100$ and $\kappa=1$. To visualise the phenomenon studied the system is perturbed here very strongly. Note the significantly higher propagation velocity of the positive density perturbation.}
\label{fig:small}
\end{center}
\end{figure}

The application of a pulse given by eq.\,(\ref{eq:field}) causes a compression of the particles on the right side and a rarefaction of the particles on the left side of the interaction region of width $w$. The positive density perturbation propagates in the $+x$ direction, while the negative density perturbation propagates in the $-x$ direction, as it will be shown later.

The propagation of these structures is traced by recording the time evolution of the spatial density distribution of the particles. To facilitate this, the simulation box is split into $M = 200$ 'stripes' along the $x$ axis and the density of the particles is measured within these stripes in each time step. Density data are saved in every 20th time step for further analysis.

The operation of the simulation method is illustrated on a small system consisting of 40\,000 particles. (The results in Sec.~\ref{sec:results} will be given for systems consisting of 4\,000\,000 particles.) For this case we set $\Gamma=100$ and $\kappa=1$, and a very high value for the perturbation field, $\widetilde{E}_0$ = 8.3 that generates density perturbations, which can easily be observed by eye on the particle snapshots. The perturbation is applied at time $t=0$ and at the position $x_0 = L/2$. 

Figure~\ref{fig:small}(a) shows snapshots of the particle positions at a time (at $\omega_{\rm p} t = - 66.7$) that belongs to the equilibrium phase, before the application of the electric field pulse at $t=0$.  Here, we find a homogeneous density distribution of the particles. Panel (b) shows a snapshot right after the perturbing pulse, at $\omega_{\rm p} t = 2.67$. A strong negative/positive density perturbation ($\delta n / n \approx 20$\%) is created left/right from the middle of the simulation box, $x/L=0.5$, as it can also be see in panel (d) of Figure~\ref{fig:small}. These perturbations propagate into opposite directions and acquire specific shapes, see e.g. panel (c) that shows a snapshot of the particle configuration at $\omega_{\rm p} t = 66.7$. At this time, the positive density peak has a sharp leading edge, which is followed by a slow decay of the density.  For the negative density peak, on the other hand, we find a slow change of the density at the leading edge and a very sharp trailing edge (also well seen in panel (d)). There is an obvious difference between the propagation velocities of the two structures, the velocity of the positive peak is about three times higher as compared to that of the negative peak. We note, that these properties are consequences of the very high degree of perturbation, for most of our studies we use significantly lower amplitude of the perturbing field, resulting in density perturbations in the order of $\delta n/n \approx$ 1\%. 


\section{Results}

\label{sec:results}

The results presented below are derived from simulations that use $N$ = 4\,000\,000 particles with a chaining mesh of $N_{\rm c}$ = 400. The cutoff distance is chosen as $r_{\rm c} = L / N_{\rm c} \cong 9 \, a$.
The width of the electric field pulse used for the generation of the solitons is set to $w \approx 35\,a$. The perturbation is applied at $x_0=L/2$ unless specified otherwise. 

In Sec.~\ref{sec:res1}, simulation results are  reported for non-magnetised systems, for different $\Gamma$ and $\kappa$ values, as well as various perturbing electric field strength, $\widetilde{E}_0$. "Collisions" of two solitons are also investigated. In Sec.~\ref{sec:res2} the effect of an external magnetic field on the propagation of the solitons in studied.

\subsection{Non-magnetized systems}

\label{sec:res1}

\begin{figure}[H]
\begin{center}
\footnotesize(a)\includegraphics[width=0.9\columnwidth]{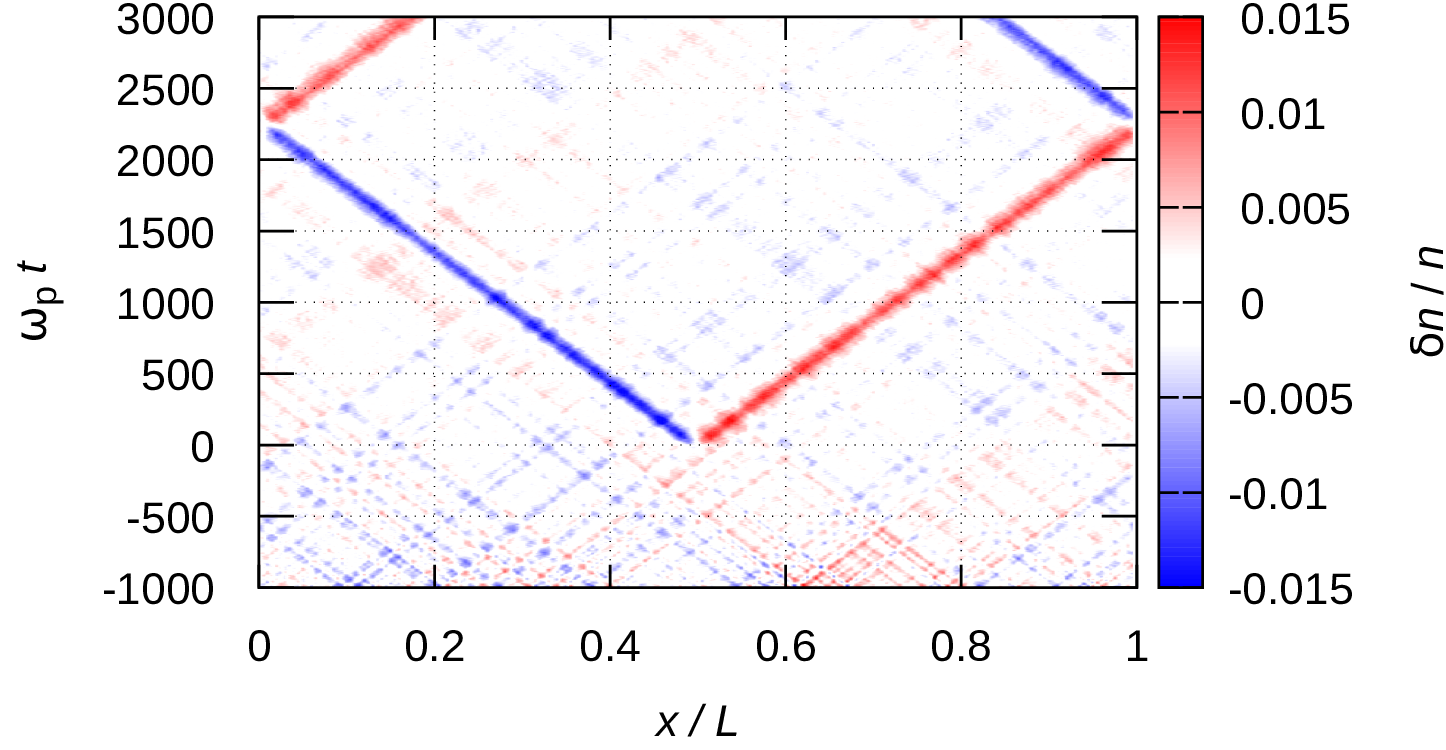}\\
\footnotesize(b)\includegraphics[width=0.9\columnwidth]{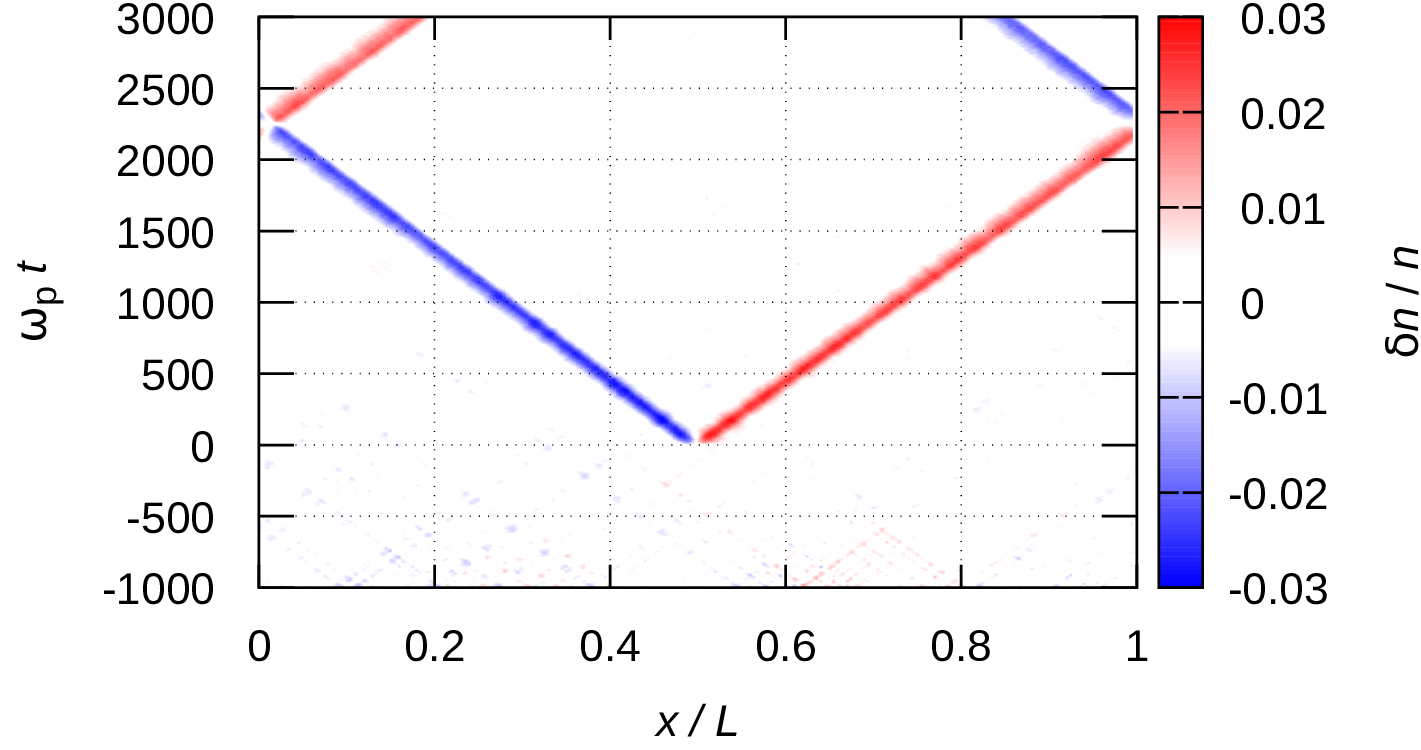}\\
\footnotesize(c)\includegraphics[width=0.9\columnwidth]{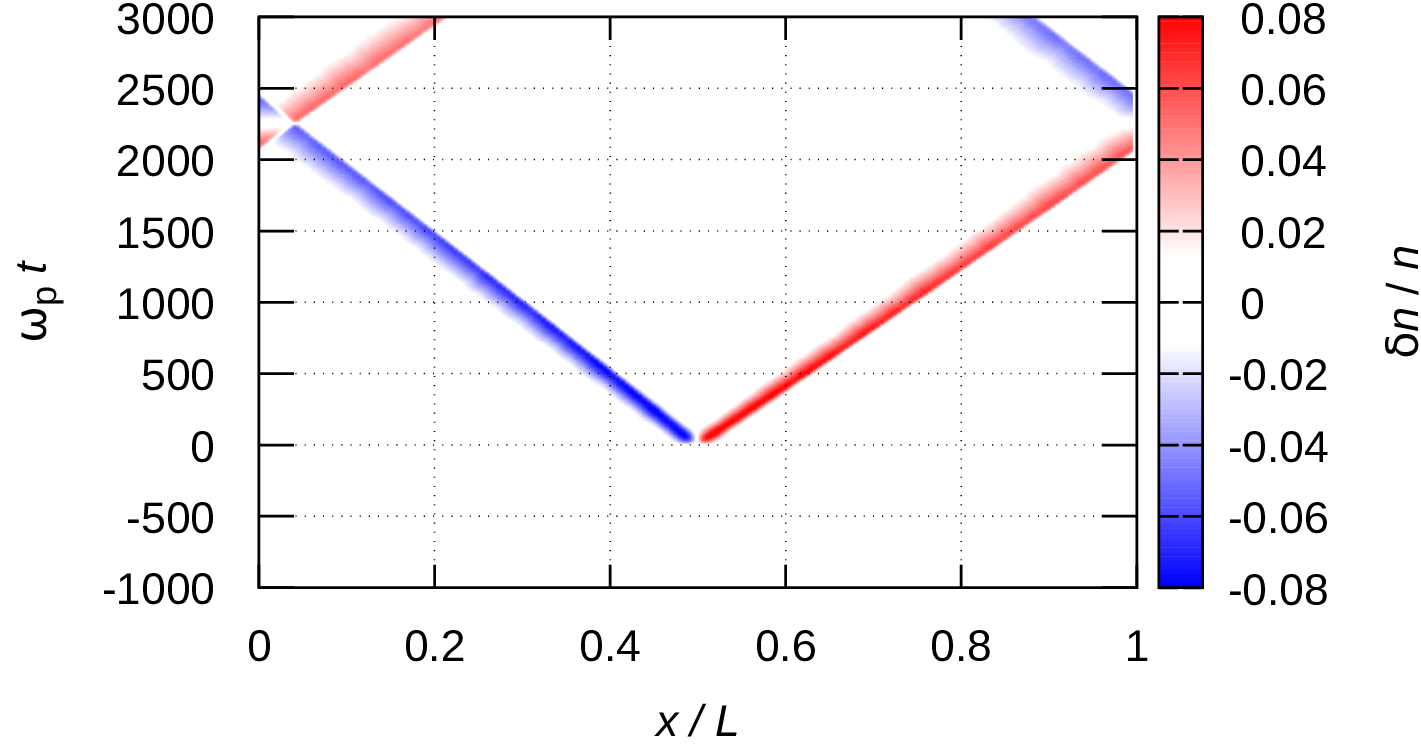}\\
\footnotesize(d)\includegraphics[width=0.9\columnwidth]{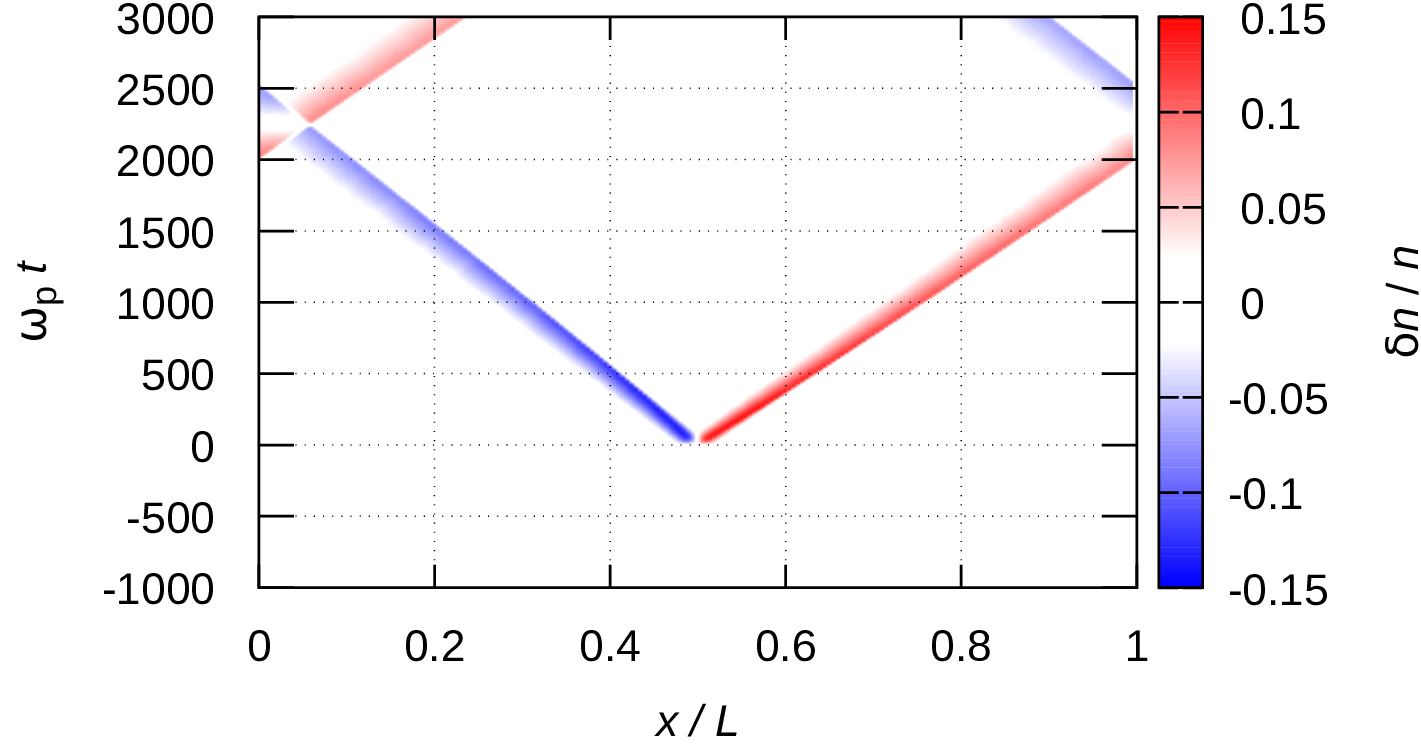}\\
\caption{Density of the system as a function of space and time. The perturbation electric field is applied at $t$ = 0, at the position $x/L = 0.5$, with amplitudes $\widetilde{E}_0$ = 0.277,  (a), 0.554 (b), 1.662 (c), and 2.77 (d).  $\Gamma=100$ and $\kappa=1$. As an effect of the periodic boundary conditions the solitons that leave the simulation box at either side, re-enter the box at the opposite side. Note, that while at low $\widetilde{E}_0$ the positive and negative density peaks propagate with the same velocity, with increasing $\widetilde{E}_0$ the velocity of the positive density peak becomes higher, and the velocity of the negative peak becomes lower. The additional features seen in (a) correspond to low-amplitude, spontaneous, propagating density fluctuations in the system.}
\label{fig:maps1}
\end{center}
\end{figure}

Figure~\ref{fig:maps1} shows the density of the particles as a function of normalized space ($x/L$) and time ($\omega_{\rm p}t$) coordinates, for different degrees of perturbation applied at $t=0$ and $x/L=0.5$. At the smallest perturbation amplitude, $\widetilde{E}_0$ = 0.277, we observe that the density changes on the scale of $\sim 1$\%. The propagating  positive and negative perturbations of the density show up as red and blue lines on this plot. The slopes of these lines are the same, i.e. both features exhibit the same velocity of propagation. At such low perturbation, low-amplitude spontaneous propagating density fluctuations are also visible to some extent. These features have the same propagation speed as the generated solitons. Similarly to the solitons, the spontaneous features have as well two branches that are created upon the initialization of the simulations. $\delta n / n$ in these two branches is, however, the same, unlike in the pairs of solitons that are created by the electric field pulse defined by eq.(\ref{eq:field}). At higher degrees of perturbation (Figures~\ref{fig:maps1}(b)--(d)) these features are no longer observed due to the broader range of $\delta n / n$ of interest. At these conditions, the propagation velocities of the "$+$" and "$-$" solitons becomes unequal: while at low $\widetilde{E}_0$ (see Figure~\ref{fig:maps1}(a)) the features "meet" at the sides of the simulation box, at higher $\widetilde{E}_0$ the positive peak propagates with a higher velocity compared to the negative peak.

\begin{figure}[H]
\begin{center}
\footnotesize(a)\includegraphics[width=0.85\columnwidth]{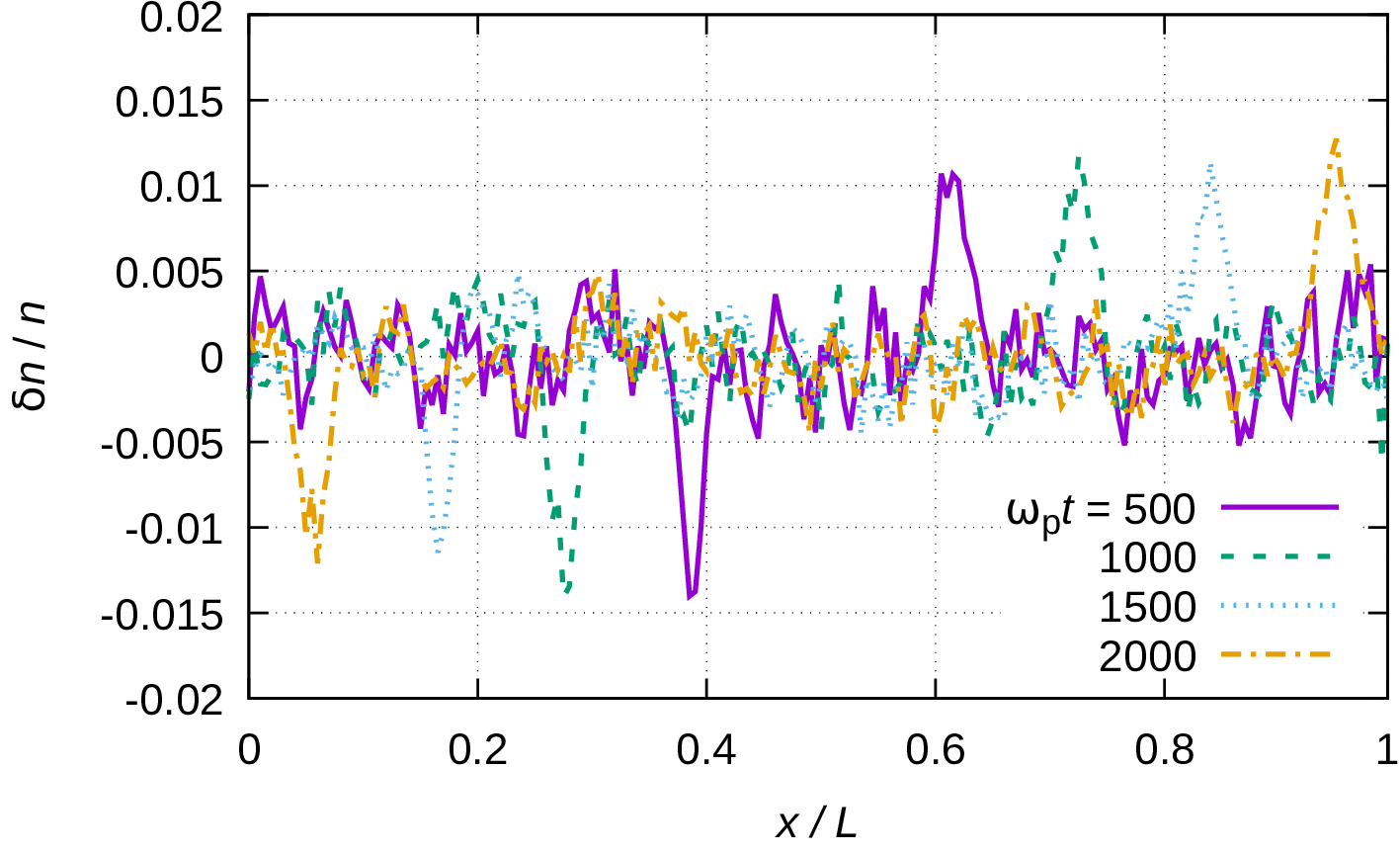}\\
\footnotesize(b)\includegraphics[width=0.85\columnwidth]{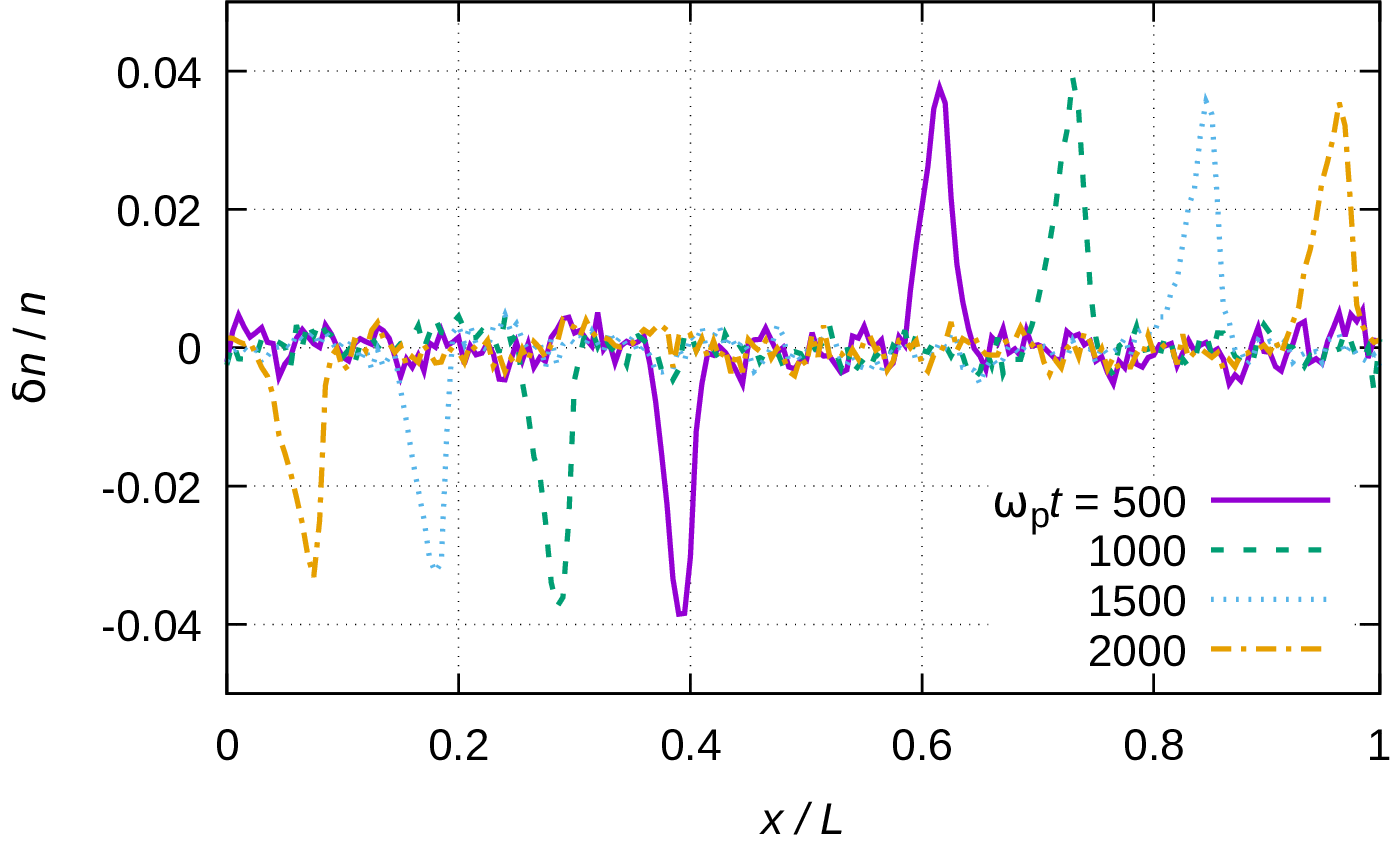}\\
\footnotesize(c)\includegraphics[width=0.85\columnwidth]{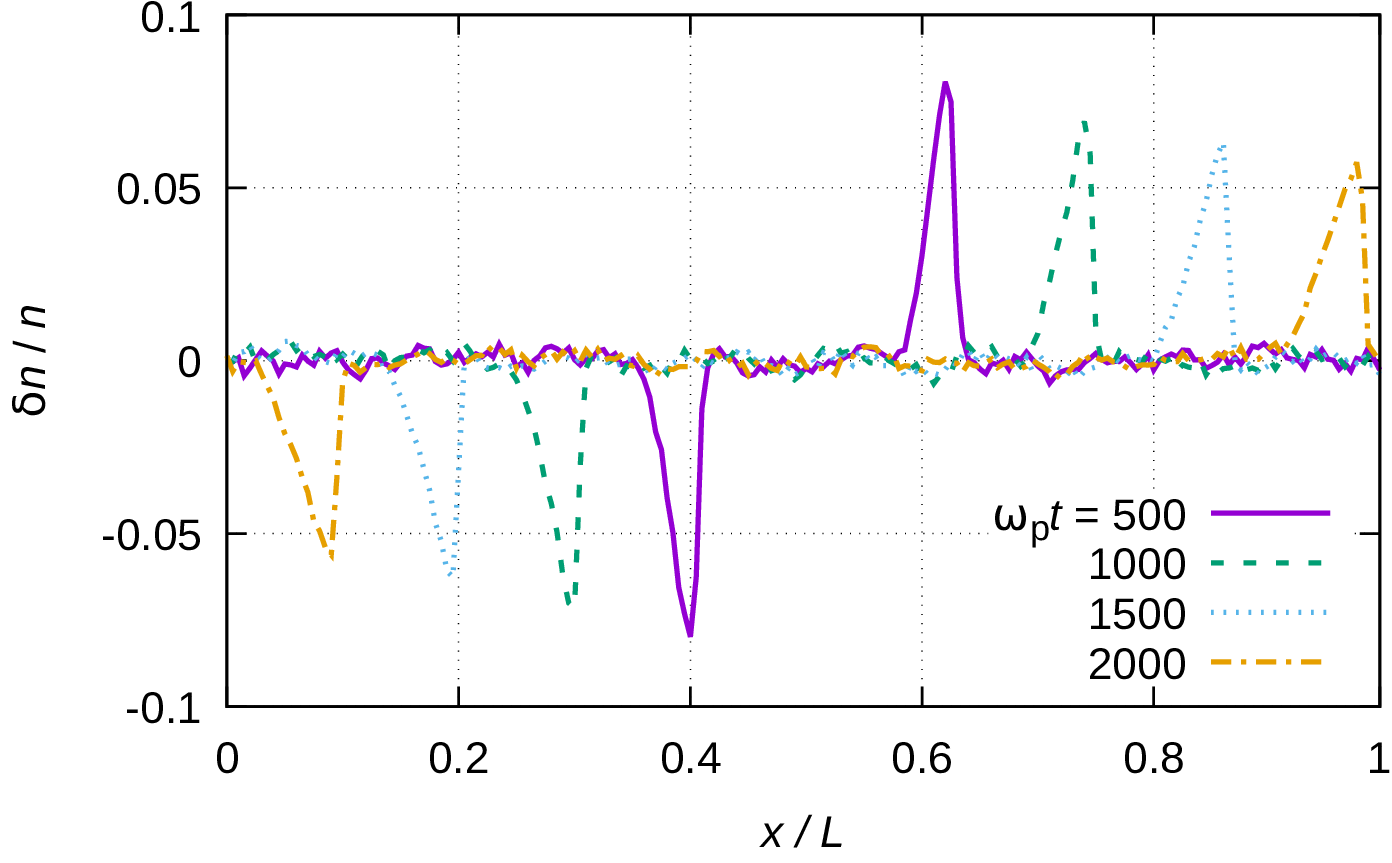}\\
\footnotesize(c)\includegraphics[width=0.85\columnwidth]{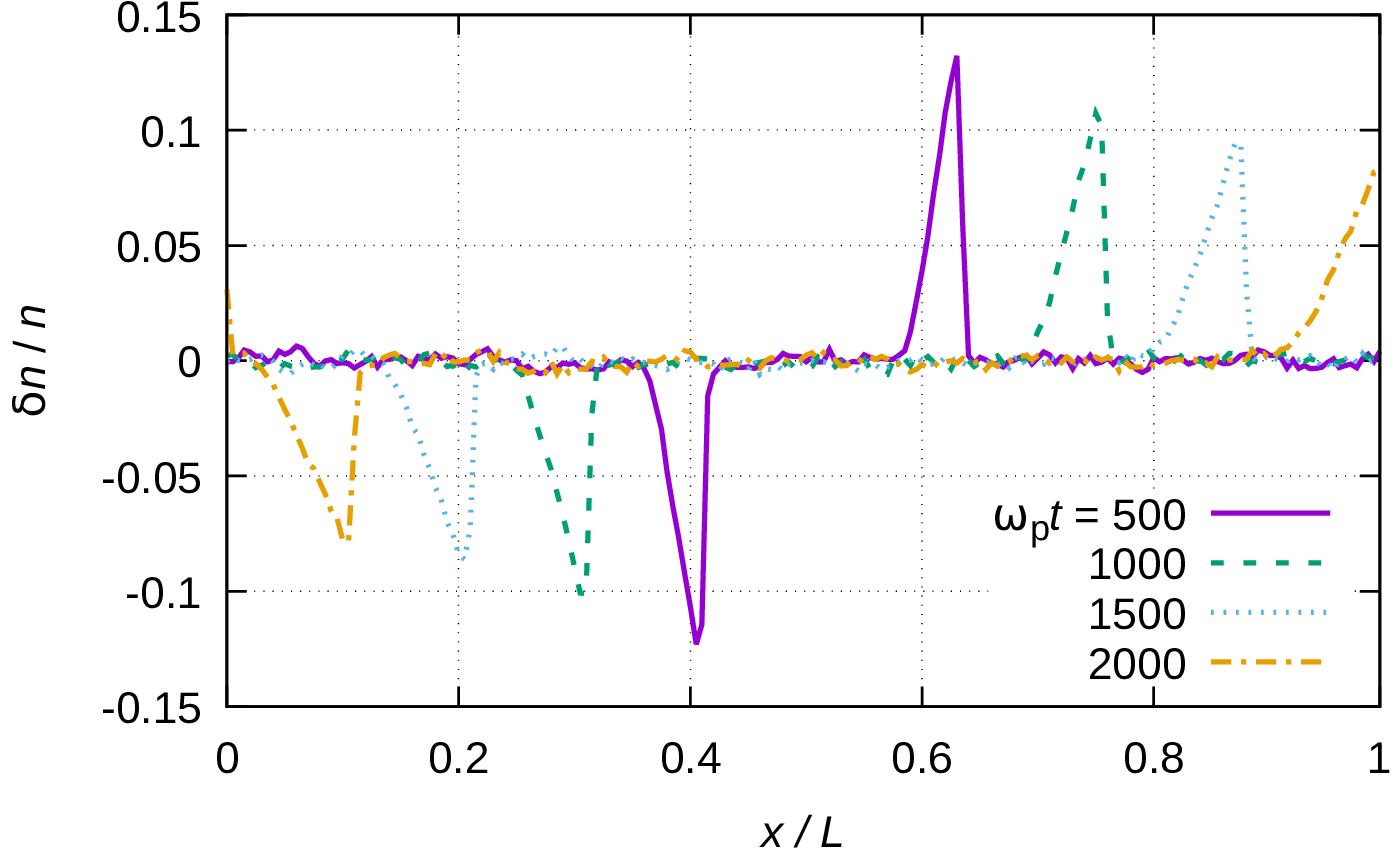}\\
\caption{Density of the system as a function of space at different times following the application of the perturbation electric field, $\widetilde{E}_0$ = 0.277 (a), 0.554 (b), 1.662 (c), and 2.77 (d). The perturbation electric field pulse is applied at $t$ = 0, at the position $x/L = 0.5$. $\Gamma=100$ and $\kappa=1$.}
\label{fig:dens1}
\end{center}
\end{figure}

The strength of the perturbation, $\widetilde{E}_0$, also influences the shapes of the propagating density perturbations. This is shown in Figure~\ref{fig:dens1}, which displays cuts of the density profiles at given times. For the case of $\widetilde{E}_0$ = 0.277 the two peaks propagate symmetrically in both directions and have similar shapes. The different propagation velocities are confirmed here, too, for the amplitudes $\widetilde{E}_0$ = 0.554, 1.662, and 2.77 (panels (b), (c), and (d), respectively). At high perturbations the shapes of both density peaks become significantly different. The positive density peak acquires a sharp leading edge and an extended trailing edge, while the opposite happens for the negative density peaks. At $\widetilde{E}_0$ = 0.277 and 0.554 (Figure~\ref{fig:dens1}(a) and (b)) the amplitude of the propagating peaks decreases only slightly with time, which is due to the broadening of the density "pulses". At higher perturbations a significant broadening as well as a remarkable change of the pulse shapes is seen in panels (c) and (d) of Figure~\ref{fig:dens1}. These effects, actually, can also be revealed from the spatio-temporal distributions shown previously in Figure~\ref{fig:maps1}. 

Our further studies are conducted with an electric field amplitude of  $\widetilde{E}_0$ = 0.554, which represents a compromise between the signal to noise ratio and small change of the density peak shapes with time. At lower $\widetilde{E}_0$ we have observed density peaks in the order of 1\% which is not much higher than the "natural" fluctuations of the density in the slabs where the density is measured. (As we use 200 slabs, the average number of particles in these is 20\,000, resulting in a fluctuation level of $\sim 1/\sqrt{20\,000} \approx 0.7$\%.) At high $\widetilde{E}_0$ values we have observed a significant change of the shapes of the density peaks over an extended domain of time.

\begin{figure}[H]
\begin{center}
\footnotesize(a)\includegraphics[width=0.45\textwidth]{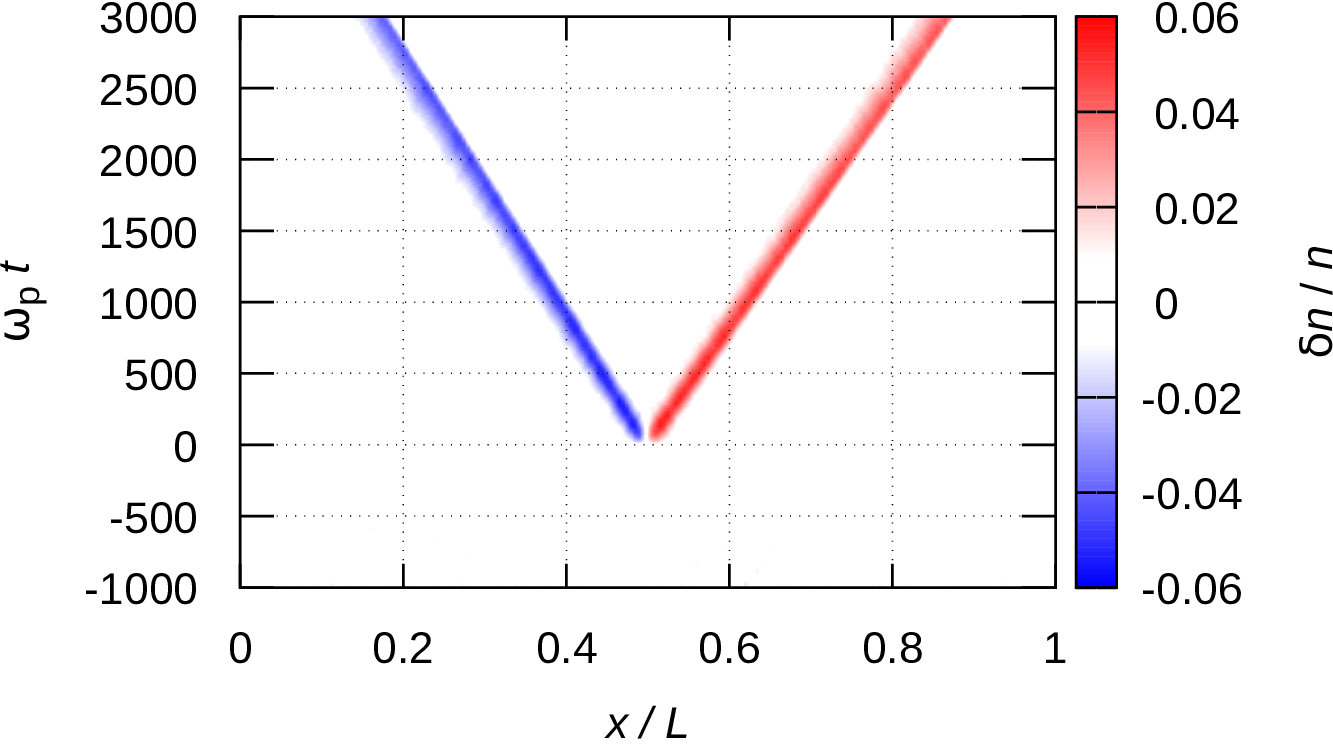}\\
\footnotesize(b)\includegraphics[width=0.45\textwidth]{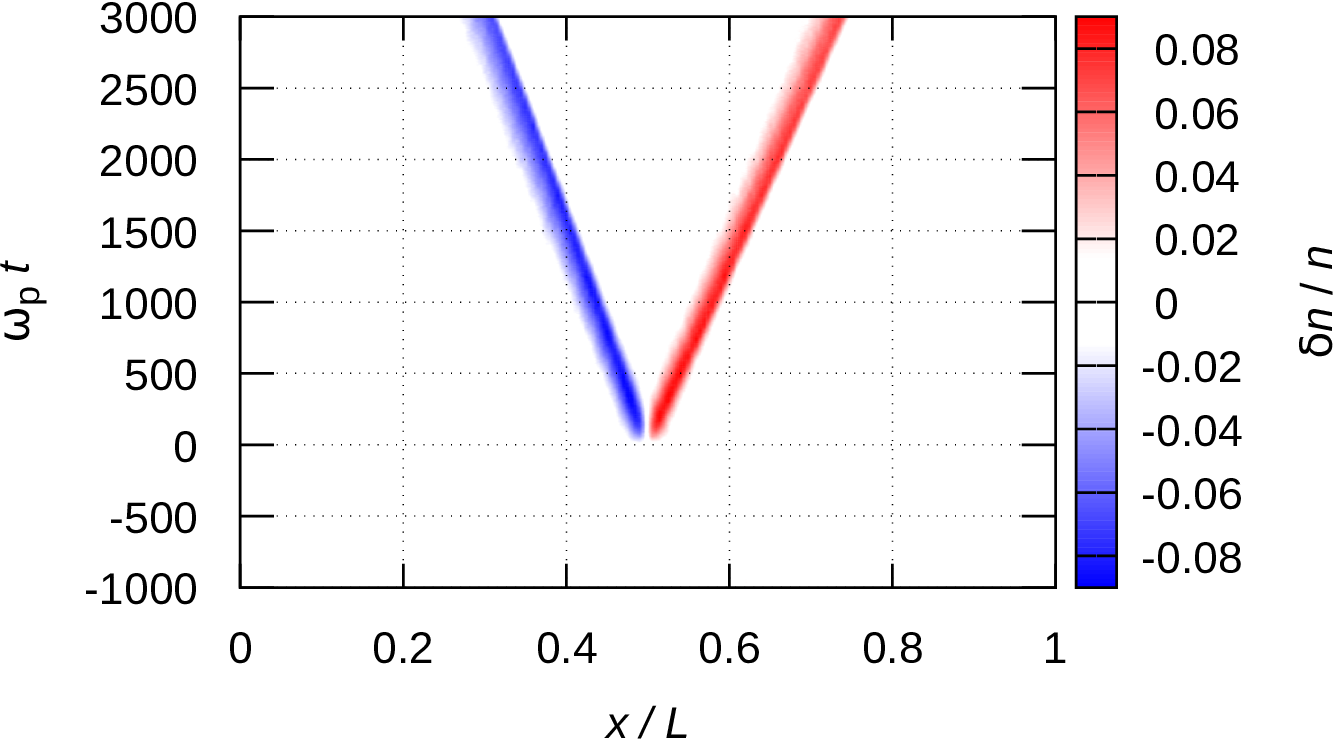}
\caption{Density of the system as a function of space and time for $\Gamma=100$ and (a) $\kappa=2$ and (b) $\kappa=3$. $\widetilde{E}_0$ = 0.554. The corresponding plot for $\kappa=1$ was shown in Figure \ref{fig:maps1}(b).}
\label{fig:maps-kappa}
\end{center}
\end{figure}

\begin{figure}[h]
\begin{center}
\footnotesize(a)\includegraphics[width=0.45\textwidth]{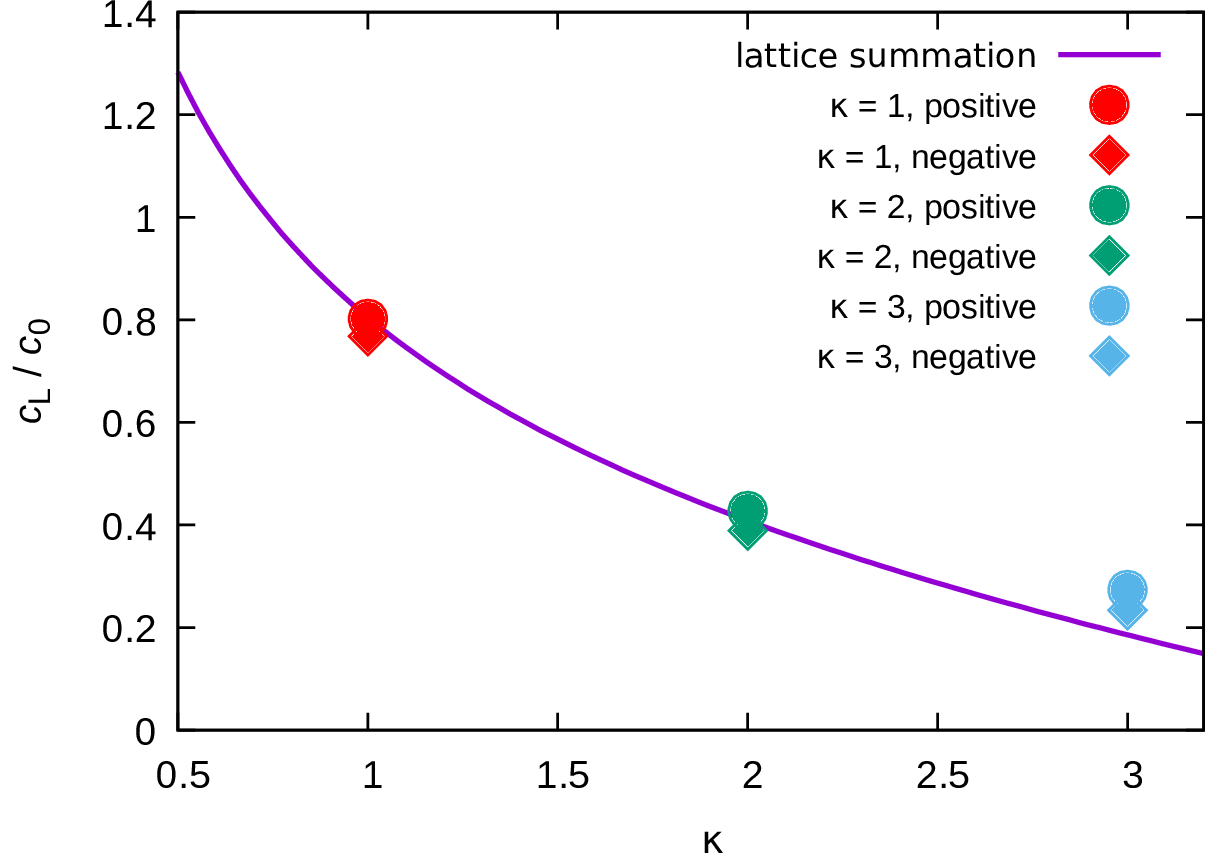}\\
\footnotesize(b)\includegraphics[width=0.45\textwidth]{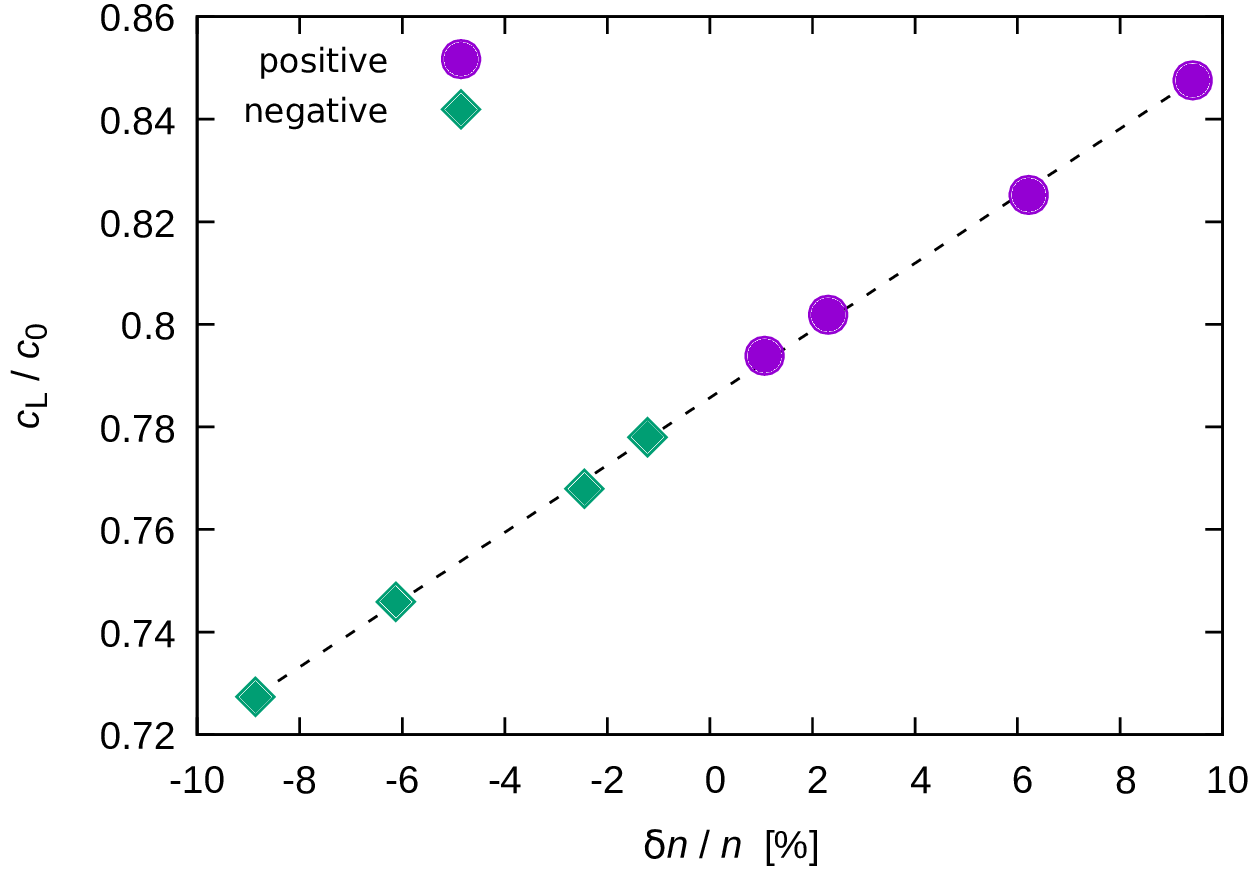}\\
\caption{Propagation velocity of the solitons: (a)  dependence on the screening parameter $\kappa$ at a fixed pulse amplitude $\widetilde{E}_0$ = 0.554 (symbols, data taken from Figure~\ref{fig:maps-kappa}). The line shows the 2D Yukawa lattice sound speed as reference (approximate formula taken from Ref. \cite{Khrapak15}). The normalization factor is $c_0 = \omega_{\rm p}a$. (b) Dependence on density modulation amplitude  including both compressional (positive) and rarefactional (negative) waves (symbols, data taken from Figure~\ref{fig:dens1}) at $\kappa=1$. The dashed line is a linear fit to the data points, shown to guide the eye. }
\label{fig:speed}
\end{center}
\end{figure}

Figure~\ref{fig:maps-kappa}, together with Figure~\ref{fig:maps1}(b) illustrates the effect of the screening parameter $\kappa$, on the propagation of the solitons. The softening of the potential (i.e., an increasing $\kappa$) clearly results in a decrease of the propagation velocity. In the limit of small amplitude solitons are found to propagate with the longitudinal sound speed. This is confirmed in Figure~\ref{fig:speed}(a) that shows the measured propagation velocities as a function of $\kappa$, in comparison of the theoretical curve derived from lattice summation calculations. The measured data are shown for both the positive and the negative density peaks, the first of these always indicates a slightly higher velocity.  Figure~\ref{fig:speed}(b) shows the propagation velocity of the solitons as a function of the density perturbation $\delta n/n$ (that in turn, depends on $\widetilde{E}_0$). The data indicate a linear dependence of the velocity on $\delta n/n$. 

\begin{figure}[h]
\begin{center}
\footnotesize(a)\includegraphics[width=0.45\textwidth]{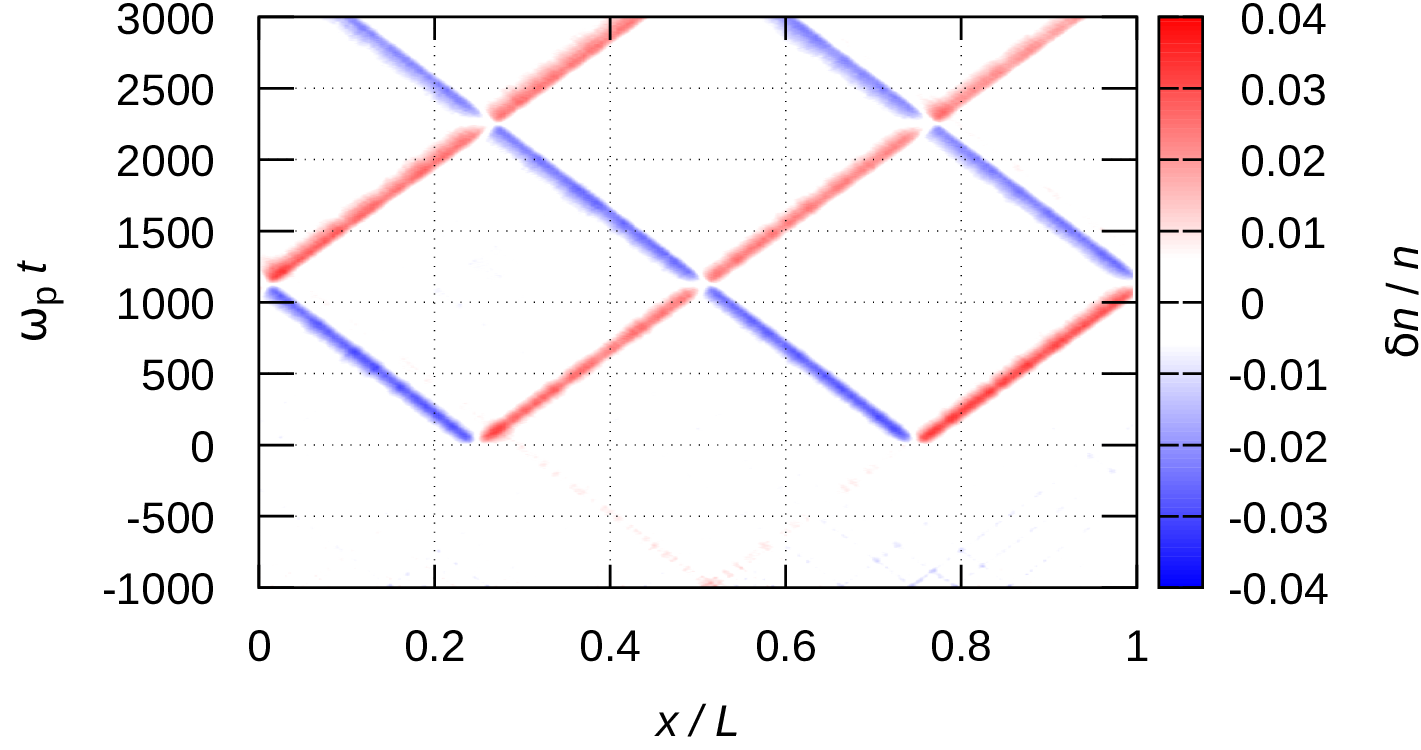}\\
\footnotesize(b)\includegraphics[width=0.45\textwidth]{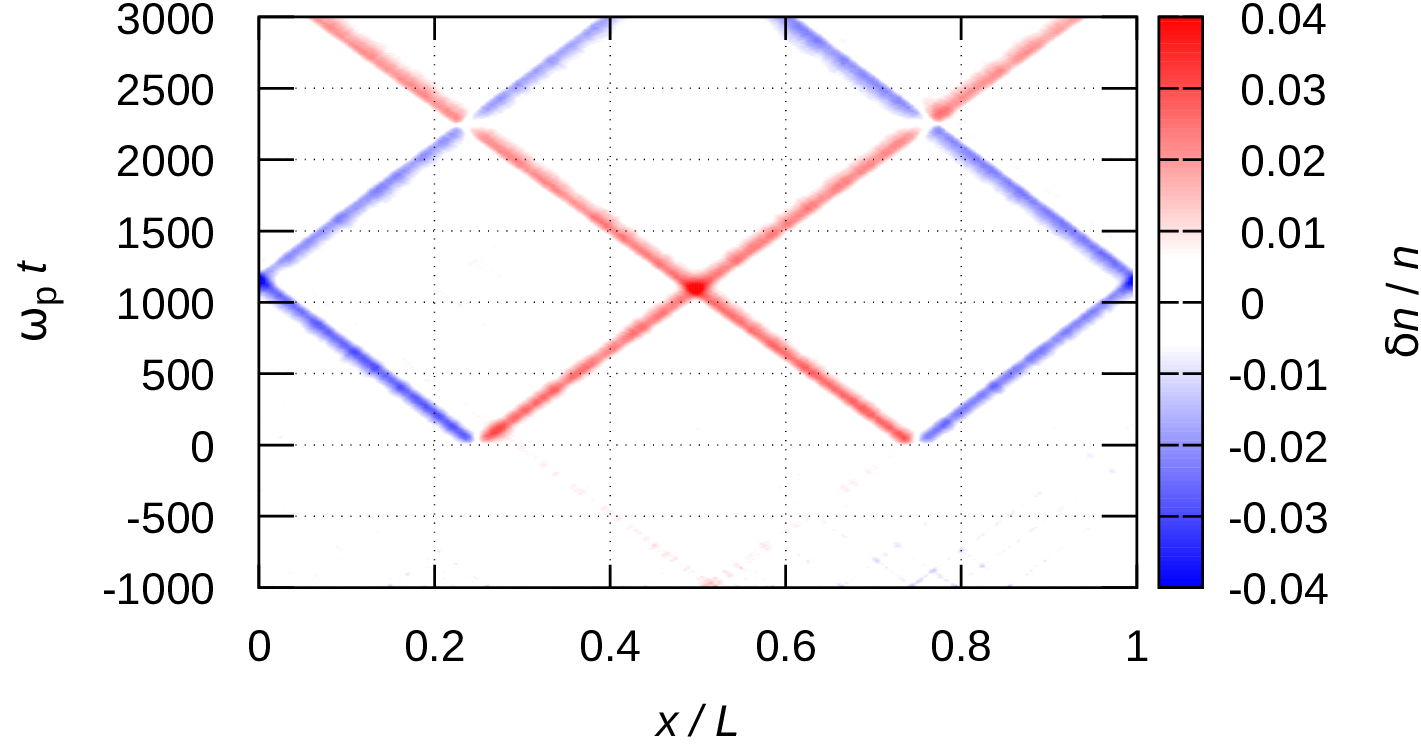}
\caption{Density of the system as a function of space and time for the cases of two pairs of solitons, generated at $x/L$ = 0.25 and at $x/L$ = 0.75. $|\widetilde{E}_0|$ = 0.554, $\Gamma=100$, $\kappa=1$. In (a) the solitons are created at the two selected spatial positions by electric field pulse having the same polarity, while in (b) the polarity of the electric field pulses is opposite at the two locations.}
\label{fig:maps_2_sol}
\end{center}
\end{figure}

Next, we investigate the scenario when the perturbing electric field is applied simultaneously at two locations in the simulation box (at $x/L$ = 0.25 and 0.75). Figure~\ref{fig:maps_2_sol}(a) shows the case when the electric field has the same polarity at the two different locations, while Figure~\ref{fig:maps_2_sol}(b) corresponds to the case when the electric field has opposite polarity at the two selected locations. In both cases, two pairs of solitons are generated. The plots of the density distributions confirm that the solitons cross each other without influencing each other's propagation. 

\subsection{The effect of the magnetic field}

\label{sec:res2}

\begin{figure}[h]
\begin{center}
\includegraphics[width=0.45\textwidth]{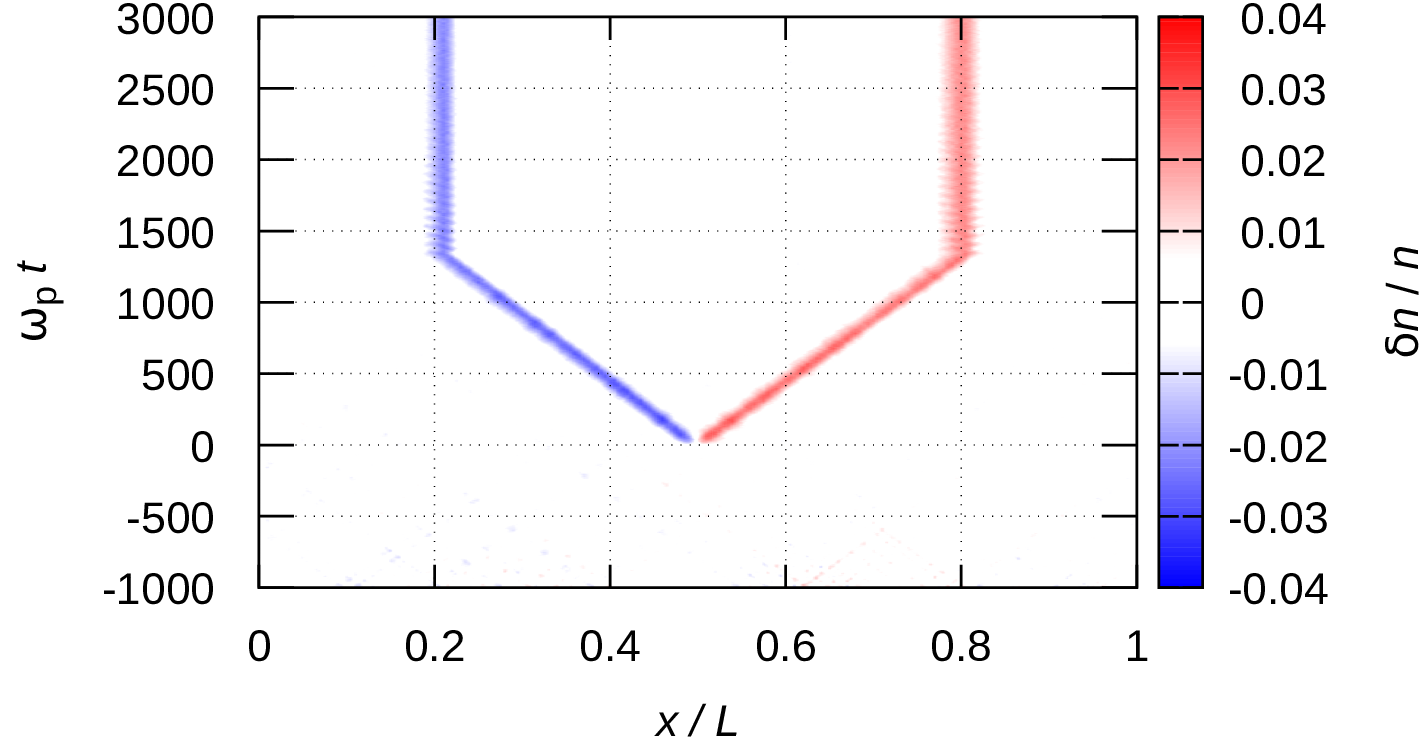}
\caption{Effect of magnetic field on the propagation of the solitons: a magnetic field with $\beta=0.1$ is turned on at $\omega_{\rm p}t = 1333.\dot{3}$. $\Gamma=100$, $\kappa=1$, $\widetilde{E}_0$ = 0.554, $x_0/L = 0.5$. Upon the application of the magnetic field the propagation of the pulses is blocked as the cyclotron motion converts propagation into localized oscillations.}
\label{fig:maps_magnetic}
\end{center}
\end{figure}

Finally, we address the  effect of an external magnetic field on the propagation of the solitons. 
In the first case, an external magnetic field with a strength of $\beta=0.1$ is turned on in the simulation at the time $\omega_{\rm p}t = 1333.\dot{3}$. Figure~\ref{fig:maps_magnetic} shows that both the positive and negative peaks become trapped, they neither propagate or dissolve by diffusion in the system over the time scale of the simulation. This behavior is the consequence of the well known scenario that diffusion can be strongly blocked by a magnetic field in strongly coupled plasmas \cite{Ott14,Ott15,Begum16,Feng17,Karasev19}. In the trapped state the particles undergo a cyclotron-type motion characterized by the strength of the magnetic field ($\beta$).

\begin{figure}[H]
\begin{center}
\footnotesize(a)\includegraphics[width=0.45\textwidth]{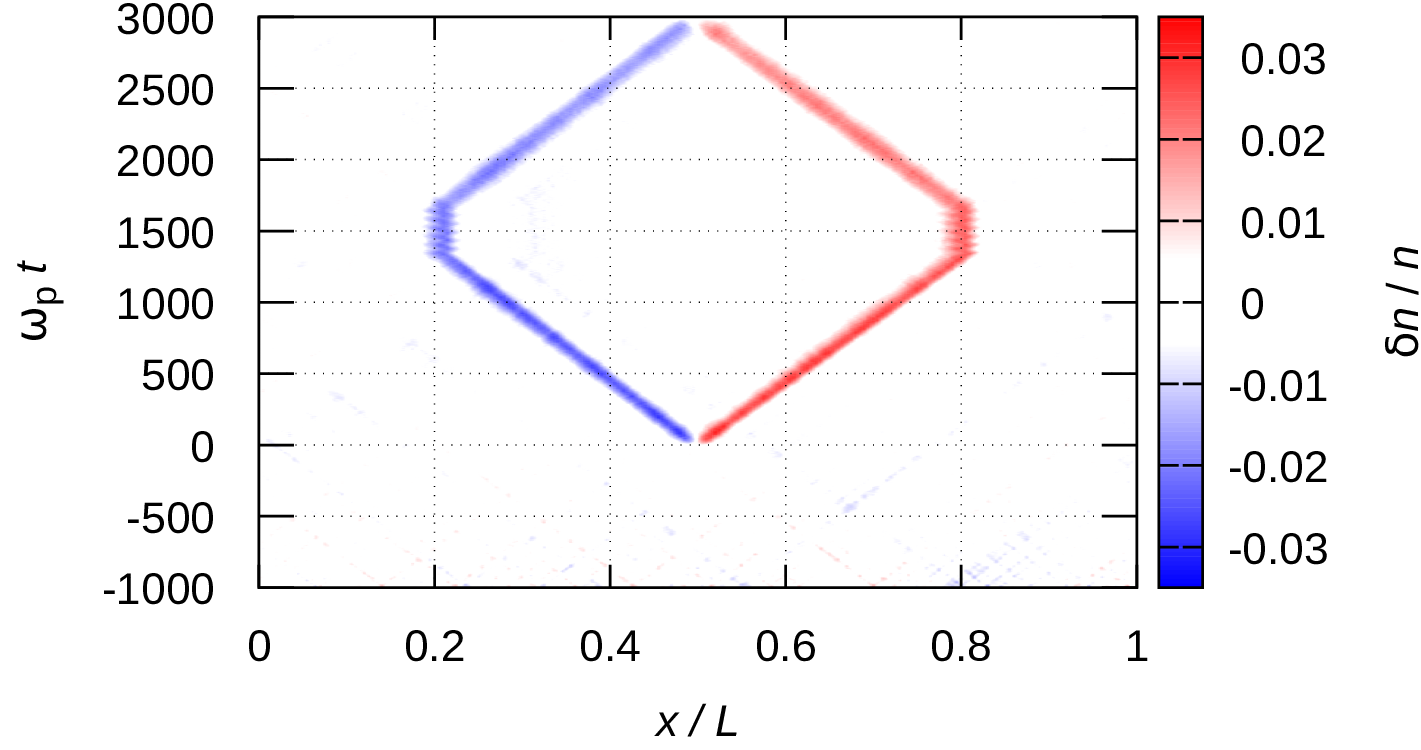}\\
\footnotesize(b)\includegraphics[width=0.45\textwidth]{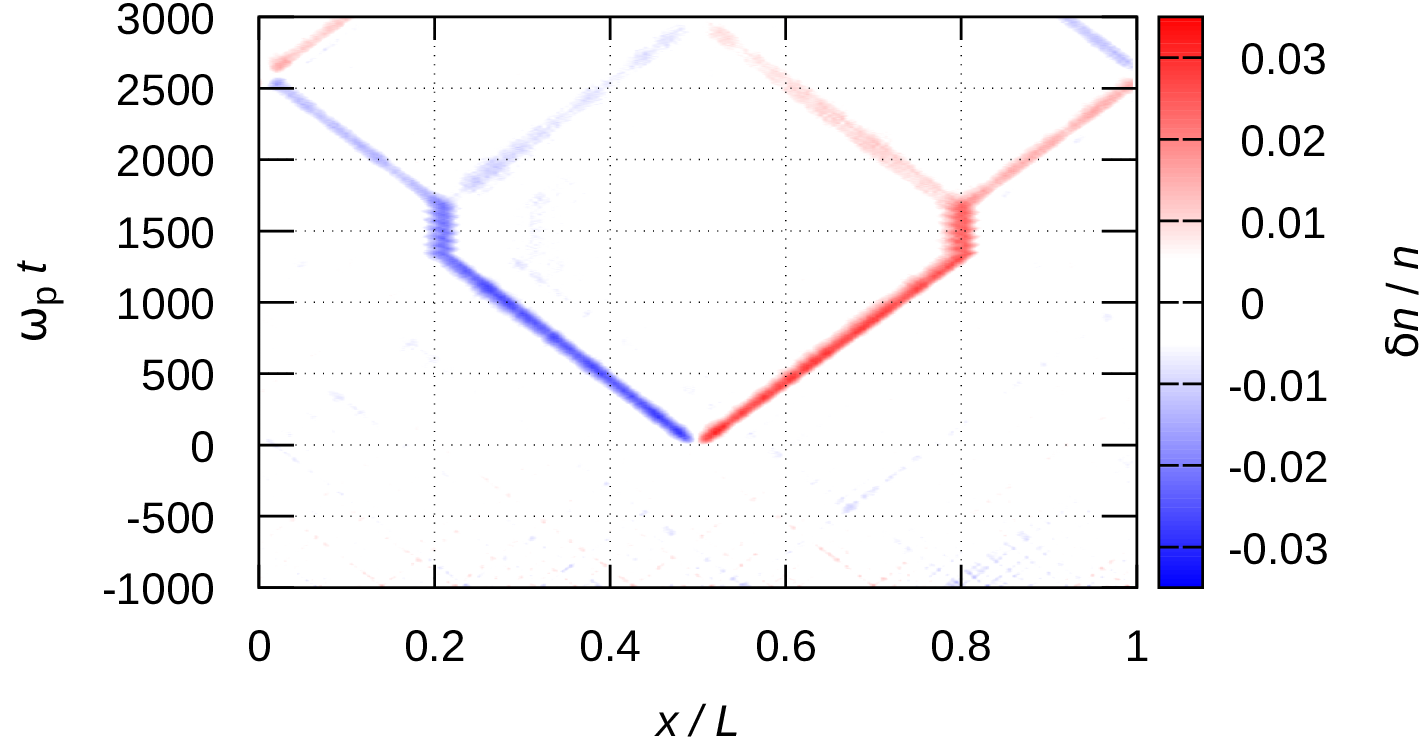}\\
\footnotesize(c)\includegraphics[width=0.45\textwidth]{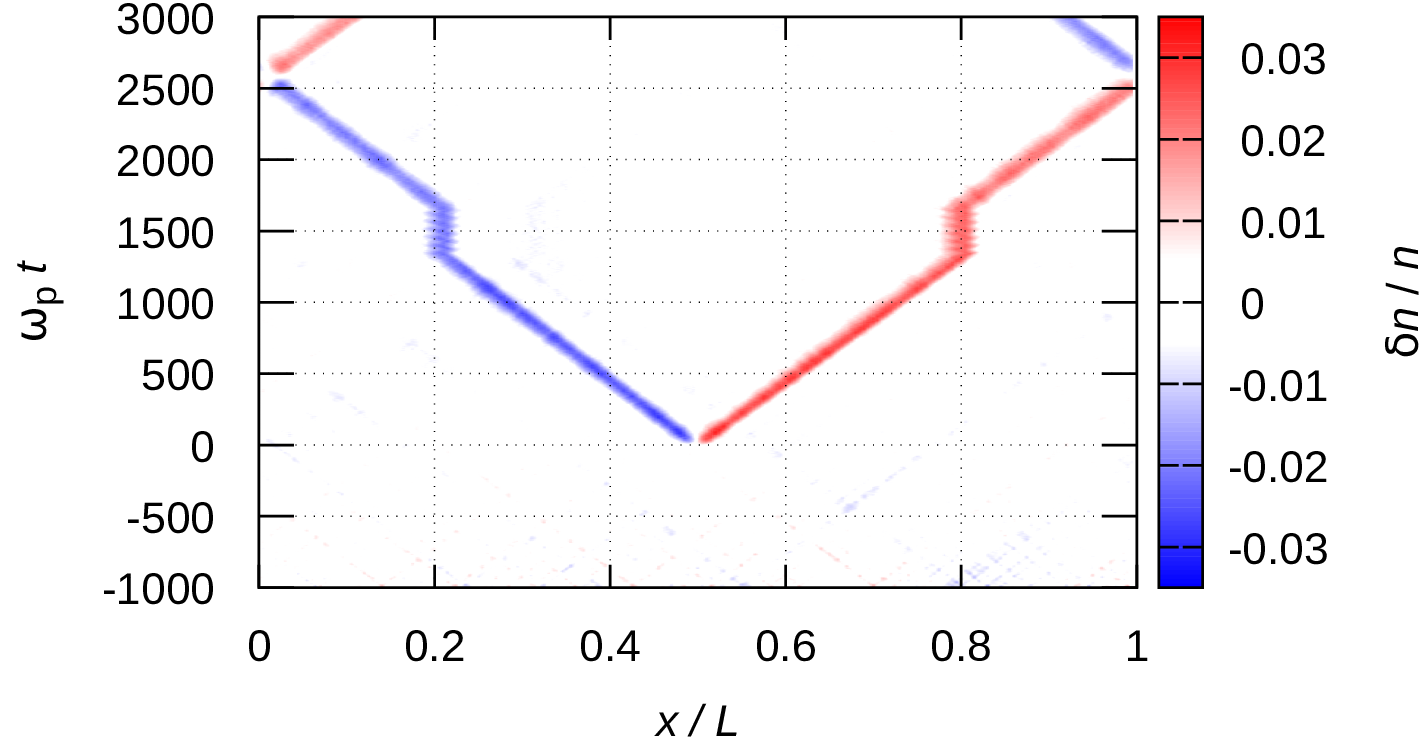}\\
\footnotesize(d)\includegraphics[width=0.45\textwidth]{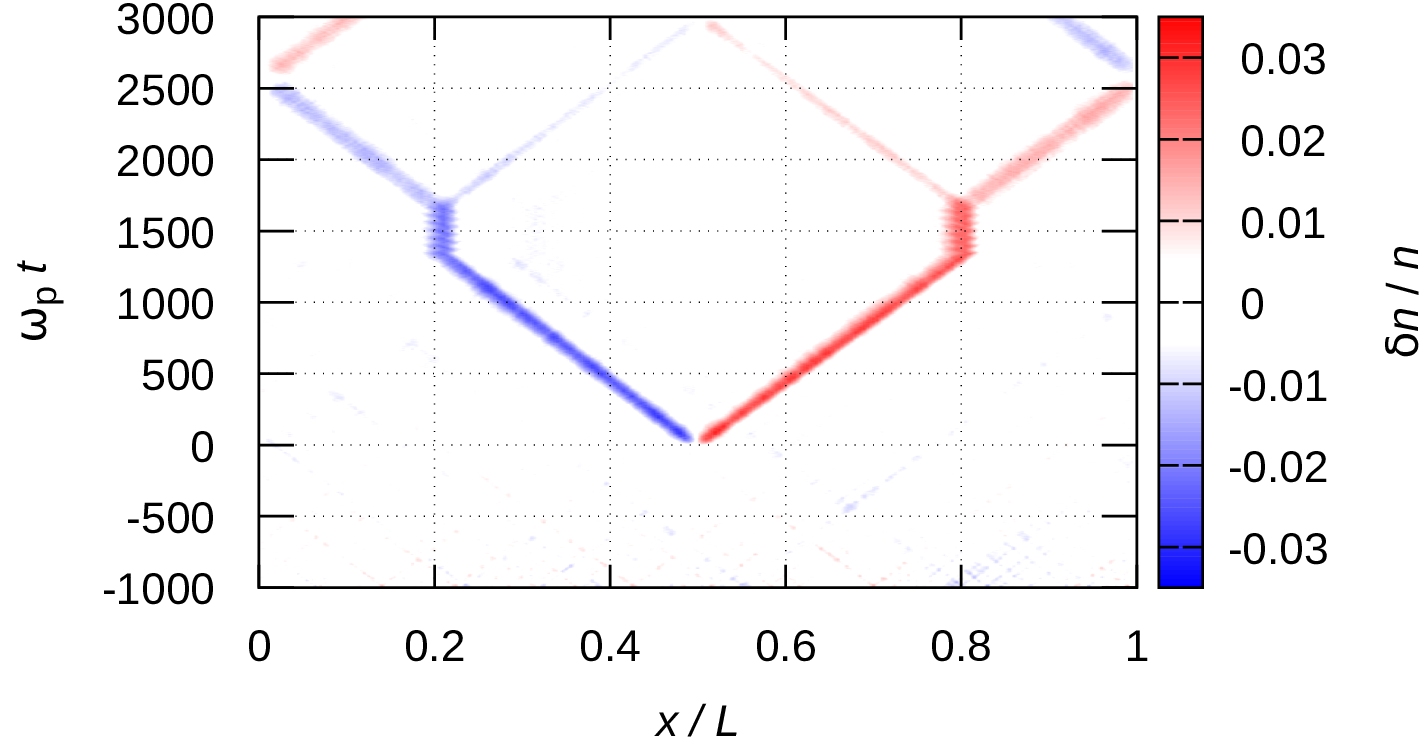}\\
\footnotesize(e)\includegraphics[width=0.45\textwidth]{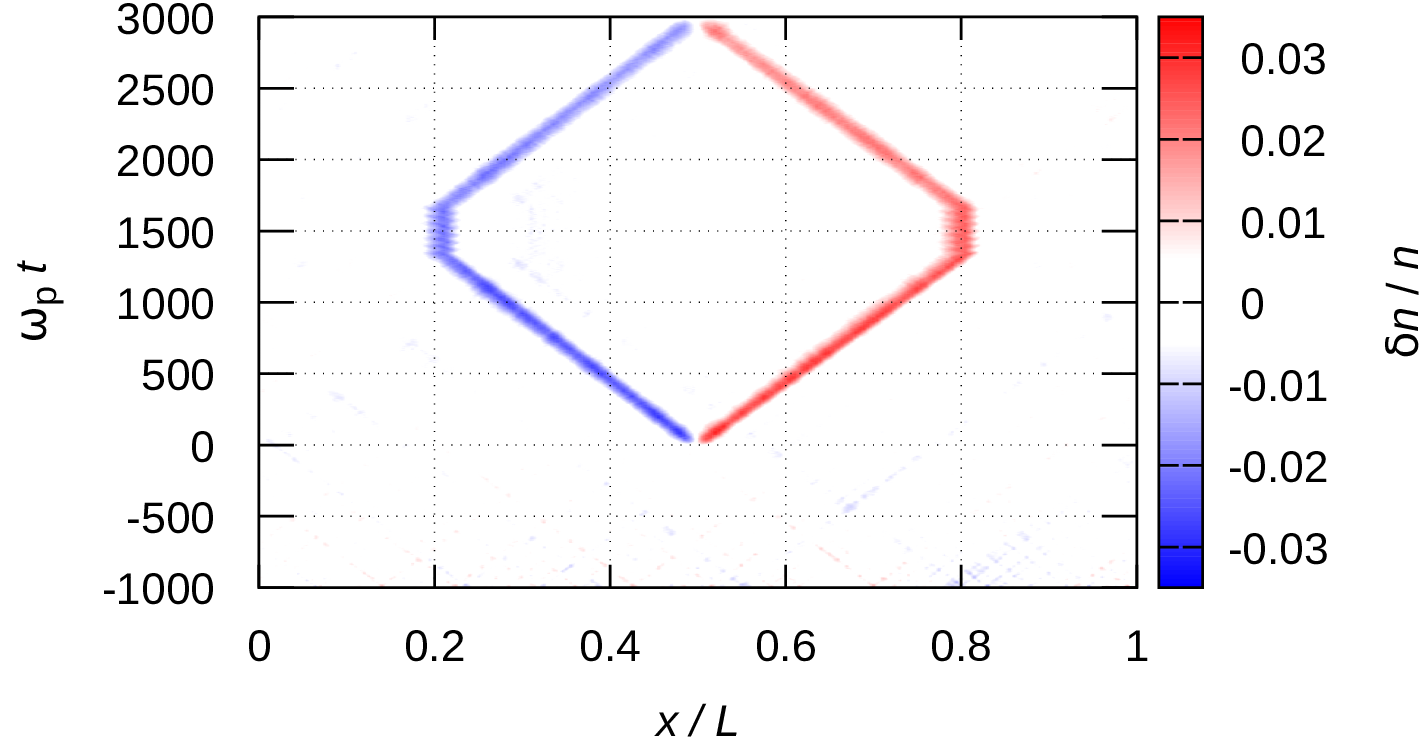}\\
\caption{Effect of a magnetic field pulse on soliton propagation: (a) $\beta=0.104$, (b) $\beta=0.108$, (c) $\beta=0.112$, (d) $\beta=0.116$, and (e) $\beta=0.120$. $\Gamma=100$, $\kappa=1$, $\widetilde{E}_0$ = 0.554. The magnetic field pulse has a duration of $\omega_{\rm p}T = 333.\dot{3}$, subsequently the solitons are released and can propagate in their original, opposite, or both  direction(s) depending on the magnetic field strength.}
\label{fig:maps_magnetic_pulse}
\end{center}
\end{figure}

\begin{table}
\caption{Parameters of cases shown in Figure 8. $\beta$ is the normalized magnetic field strength, $N_{\rm c} = \Omega_{\rm c}\,T\,/\, 2\,\pi$ is the number of cyclotron oscillation cycles during the on phase of the magnetic field, $\phi$ is the phase angle of the fractional part of $N_{\rm c}$.}
\begin{ruledtabular}
\begin{tabular}{ccccc}
$\beta$ & $N_{\rm c}$ & $\phi$\,[deg] & Effect & Figure \\
\hline
0.102   & 5.411  & 148.1 & Reflection  & \\
0.104   & 5.517  & 186.3 & Reflection   & 8(a) \\
0.106   & 5.623  & 224.4 & Splitting & \\
0.108   & 5.730  & 262.6 & Splitting   & 8(b) \\
0.110   & 5.836  & 300.8 & Transmission  &  \\
0.112   & 5.942  & 339.0 & Transmission  & 8(c) \\
0.114   & 6.048  & 17.2 & Transmission   &  \\
0.116   & 6.154  & 55.4 & Splitting & 8(d) \\
0.118   & 6.260  & 93.6 & Splitting   &  \\
0.120   & 6.366  & 131.8 & Reflection  & 8(e) \\
0.122   & 6.472  & 170.0 & Reflection  &  \\
0.124   & 6.578  & 208.2 & Reflection  &  \\
\end{tabular}
\end{ruledtabular}
\label{tab:1}
\end{table}

It is interesting to note, however, that if the magnetic field is switched off after a certain time, the temporarily blocked solitons can be "released". What happens at this moment is defined by the phase of the cyclotron motion. Figure~\ref{fig:maps_magnetic_pulse} displays and Table~\ref{tab:1} lists a sequence of cases with small differences in the magnetic field strength ($\beta$).

Depending on the phase of the cyclotron oscillation of the trapped particles the solitons can (i) continue  propagating into their original directions, termed as "transmission" in Table I, (ii) propagate into the opposite directions, termed as "reflection" in Table I, as well as (iii) split into a pair of solitons having the same polarity, termed as "splitting" in Table I.  

For the cases studied the on phase of the magnetic field pulse has a fixed duration of $\omega_{\rm p}\,T = 333.\dot{3}$. During this time the particles undergo a number of cyclotron oscillations $N_{\rm c} = \Omega_{\rm c} T / 2 \pi = 333.\dot{3}\, (\Omega_{\rm c}/\omega_{\rm p}) / 2\pi = 333.\dot{3}\, \beta / 2\pi$. Along with the values of $\beta$, Table I. also gives the valus of $N_{\rm c}$ as defined above and its fractional part converted to a phase angle $0^\circ \leq \phi \leq 360^\circ$. Our data show that whenever $\phi$ is close to $0^\circ$ or $360^\circ$, the solitons are transmitted after the temporary trapping by the magnetic field pulse. Reflection occurs whenever $\phi$ is in the vicinity of $180^\circ$, as expected because of the delay of phase of the localized cyclotron oscillations. At intermediate values of $\phi$, close to $90^\circ$ and $270^\circ$ splitting occurs with nearly equal or different amplitudes depending on the exact value of $\phi$. It is expcted that the localization time (duration of the magnetic field pulse) has a similar effect as the effect of the magnetic field strength, as $N_{\rm c}$, and consequently $\phi$ are proportional to $T$.

\section{Summary}

\label{sec:summary}

This work reported our investigations on the propagation of solitons, created by electric field pulses, in a two-dimensional strongly-coupled many body system characterized by the Yukawa potential. The electric field pulses created a pair of solitons, with positive / negative density peaks (referenced to the density of the unperturbed system) that were found to propagate into opposite directions. The propagation speed of both features in the limit of small density perturbations was found to be equal to the longitudinal sound speed in the Yukawa liquid. With increasing perturbation the propagation speed of the positive peak was found to increase, while the propagation speed of the negative peak was found to decrease. 

An external magnetic field was found to "freeze" the positions of the density peaks (solitons) due to the largely reduced self-diffusion in the system at $\beta > 0$. Upon the termination of this magnetic field, however, the solitons were found to be released from these "traps" and to continue propagating into directions that depends on the strength of the magnetic field and the trapping time. These observations call for further studies at the microscopic level of individual particles and for the exploration of multiple solutions trapped simultaneously by magnetic field pulses.

\acknowledgements 

This work has been supported by the Hungarian Office for Research, Development, and Innovation (NKFIH 119357 and 115805) and by the grant AP05132665 of the Ministry of Education and Science of the Republic of Kazakhstan.

\bibliography{solitonbib}

\begin{thebibliography}{71}%
\makeatletter
\providecommand \@ifxundefined [1]{%
 \@ifx{#1\undefined}
}%
\providecommand \@ifnum [1]{%
 \ifnum #1\expandafter \@firstoftwo
 \else \expandafter \@secondoftwo
 \fi
}%
\providecommand \@ifx [1]{%
 \ifx #1\expandafter \@firstoftwo
 \else \expandafter \@secondoftwo
 \fi
}%
\providecommand \natexlab [1]{#1}%
\providecommand \enquote  [1]{``#1''}%
\providecommand \bibnamefont  [1]{#1}%
\providecommand \bibfnamefont [1]{#1}%
\providecommand \citenamefont [1]{#1}%
\providecommand \href@noop [0]{\@secondoftwo}%
\providecommand \href [0]{\begingroup \@sanitize@url \@href}%
\providecommand \@href[1]{\@@startlink{#1}\@@href}%
\providecommand \@@href[1]{\endgroup#1\@@endlink}%
\providecommand \@sanitize@url [0]{\catcode `\\12\catcode `\$12\catcode
  `\&12\catcode `\#12\catcode `\^12\catcode `\_12\catcode `\%12\relax}%
\providecommand \@@startlink[1]{}%
\providecommand \@@endlink[0]{}%
\providecommand \url  [0]{\begingroup\@sanitize@url \@url }%
\providecommand \@url [1]{\endgroup\@href {#1}{\urlprefix }}%
\providecommand \urlprefix  [0]{URL }%
\providecommand \Eprint [0]{\href }%
\providecommand \doibase [0]{http://dx.doi.org/}%
\providecommand \selectlanguage [0]{\@gobble}%
\providecommand \bibinfo  [0]{\@secondoftwo}%
\providecommand \bibfield  [0]{\@secondoftwo}%
\providecommand \translation [1]{[#1]}%
\providecommand \BibitemOpen [0]{}%
\providecommand \bibitemStop [0]{}%
\providecommand \bibitemNoStop [0]{.\EOS\space}%
\providecommand \EOS [0]{\spacefactor3000\relax}%
\providecommand \BibitemShut  [1]{\csname bibitem#1\endcsname}%
\let\auto@bib@innerbib\@empty
\bibitem [{\citenamefont {Robinson}\ and\ \citenamefont
  {Russel}(1838)}]{Russel1838}%
  \BibitemOpen
  \bibfield  {author} {\bibinfo {author} {\bibfnamefont {J.}~\bibnamefont
  {Robinson}}\ and\ \bibinfo {author} {\bibfnamefont {J.}~\bibnamefont
  {Russel}},\ }\href@noop {} {\emph {\bibinfo {title} {in Report of the
  Committee on Waves, Report of the 7th Meeting of British Association for the
  Advancement of Science}}}\ (\bibinfo  {publisher} {John Murray, London,
  Liverpool},\ \bibinfo {year} {1838})\ p.\ \bibinfo {pages} {417}\BibitemShut
  {NoStop}%
\bibitem [{\citenamefont {Russell}(2018)}]{Russel1845}%
  \BibitemOpen
  \bibfield  {author} {\bibinfo {author} {\bibfnamefont {J.~S.}\ \bibnamefont
  {Russell}},\ }\href {https://books.google.hu/books?id=jggOtQEACAAJ} {\emph
  {\bibinfo {title} {Report on Waves: Made to the Meetings of the British
  Association in 1842-43}}}\ (\bibinfo  {publisher} {Creative Media Partners,
  LLC},\ \bibinfo {year} {1845 and 2018})\BibitemShut {NoStop}%
\bibitem [{\citenamefont {Korteweg}\ and\ \citenamefont
  {de~Vries}(1895)}]{Korteweg1896}%
  \BibitemOpen
  \bibfield  {author} {\bibinfo {author} {\bibfnamefont {D.~D.~J.}\
  \bibnamefont {Korteweg}}\ and\ \bibinfo {author} {\bibfnamefont {D.~G.}\
  \bibnamefont {de~Vries}},\ }\href {\doibase 10.1080/14786449508620739}
  {\bibfield  {journal} {\bibinfo  {journal} {The London, Edinburgh, and Dublin
  Philosophical Magazine and Journal of Science}\ }\textbf {\bibinfo {volume}
  {39}},\ \bibinfo {pages} {422} (\bibinfo {year} {1895})}\BibitemShut
  {NoStop}%
\bibitem [{\citenamefont {Hasegawa}\ and\ \citenamefont
  {Tappert}(1973)}]{Hasegawa73}%
  \BibitemOpen
  \bibfield  {author} {\bibinfo {author} {\bibfnamefont {A.}~\bibnamefont
  {Hasegawa}}\ and\ \bibinfo {author} {\bibfnamefont {F.}~\bibnamefont
  {Tappert}},\ }\href@noop {} {\bibfield  {journal} {\bibinfo  {journal} {Appl.
  Phys. Lett.}\ }\textbf {\bibinfo {volume} {23}},\ \bibinfo {pages} {142}
  (\bibinfo {year} {1973})}\BibitemShut {NoStop}%
\bibitem [{\citenamefont {Kosevich}\ \emph {et~al.}(1998)\citenamefont
  {Kosevich}, \citenamefont {Gann}, \citenamefont {Zhukov},\ and\ \citenamefont
  {Voronov}}]{Kosevich1998}%
  \BibitemOpen
  \bibfield  {author} {\bibinfo {author} {\bibfnamefont {A.~M.}\ \bibnamefont
  {Kosevich}}, \bibinfo {author} {\bibfnamefont {V.~V.}\ \bibnamefont {Gann}},
  \bibinfo {author} {\bibfnamefont {A.~I.}\ \bibnamefont {Zhukov}}, \ and\
  \bibinfo {author} {\bibfnamefont {V.~P.}\ \bibnamefont {Voronov}},\ }\href
  {\doibase 10.1134/1.558674} {\bibfield  {journal} {\bibinfo  {journal}
  {Journal of Experimental and Theoretical Physics}\ }\textbf {\bibinfo
  {volume} {87}},\ \bibinfo {pages} {401} (\bibinfo {year} {1998})}\BibitemShut
  {NoStop}%
\bibitem [{\citenamefont {Iwata}\ and\ \citenamefont
  {Stevenson}(2019)}]{Iwata19}%
  \BibitemOpen
  \bibfield  {author} {\bibinfo {author} {\bibfnamefont {Y.}~\bibnamefont
  {Iwata}}\ and\ \bibinfo {author} {\bibfnamefont {P.}~\bibnamefont
  {Stevenson}},\ }\href {\doibase 10.1088/1367-2630/ab0e58} {\bibfield
  {journal} {\bibinfo  {journal} {New Journal of Physics}\ }\textbf {\bibinfo
  {volume} {21}},\ \bibinfo {pages} {043010} (\bibinfo {year}
  {2019})}\BibitemShut {NoStop}%
\bibitem [{\citenamefont {Frantzeskakis}(2010)}]{Frantzeskakis10}%
  \BibitemOpen
  \bibfield  {author} {\bibinfo {author} {\bibfnamefont {D.~J.}\ \bibnamefont
  {Frantzeskakis}},\ }\href {\doibase 10.1088/1751-8113/43/21/213001}
  {\bibfield  {journal} {\bibinfo  {journal} {Journal of Physics A:
  Mathematical and Theoretical}\ }\textbf {\bibinfo {volume} {43}},\ \bibinfo
  {pages} {213001} (\bibinfo {year} {2010})}\BibitemShut {NoStop}%
\bibitem [{\citenamefont {Tran}(1979)}]{Tran79}%
  \BibitemOpen
  \bibfield  {author} {\bibinfo {author} {\bibfnamefont {M.~Q.}\ \bibnamefont
  {Tran}},\ }\href {\doibase 10.1088/0031-8949/20/3-4/004} {\bibfield
  {journal} {\bibinfo  {journal} {Physica Scripta}\ }\textbf {\bibinfo {volume}
  {20}},\ \bibinfo {pages} {317} (\bibinfo {year} {1979})}\BibitemShut
  {NoStop}%
\bibitem [{\citenamefont {Olivier}\ \emph {et~al.}(2018)\citenamefont
  {Olivier}, \citenamefont {Verheest},\ and\ \citenamefont
  {Hereman}}]{Olivier18}%
  \BibitemOpen
  \bibfield  {author} {\bibinfo {author} {\bibfnamefont {C.~P.}\ \bibnamefont
  {Olivier}}, \bibinfo {author} {\bibfnamefont {F.}~\bibnamefont {Verheest}}, \
  and\ \bibinfo {author} {\bibfnamefont {W.~A.}\ \bibnamefont {Hereman}},\
  }\href {\doibase 10.1063/1.5027448} {\bibfield  {journal} {\bibinfo
  {journal} {Physics of Plasmas}\ }\textbf {\bibinfo {volume} {25}},\ \bibinfo
  {pages} {032309} (\bibinfo {year} {2018})}\BibitemShut {NoStop}%
\bibitem [{\citenamefont {Verheest}\ and\ \citenamefont
  {Hereman}(2019)}]{Verheest19}%
  \BibitemOpen
  \bibfield  {author} {\bibinfo {author} {\bibfnamefont {F.}~\bibnamefont
  {Verheest}}\ and\ \bibinfo {author} {\bibfnamefont {W.~A.}\ \bibnamefont
  {Hereman}},\ }\href {\doibase 10.1017/S0022377818001368} {\bibfield
  {journal} {\bibinfo  {journal} {Journal of Plasma Physics}\ }\textbf
  {\bibinfo {volume} {85}},\ \bibinfo {pages} {905850106} (\bibinfo {year}
  {2019})}\BibitemShut {NoStop}%
\bibitem [{\citenamefont {EL-Shamy}\ \emph {et~al.}(2019)\citenamefont
  {EL-Shamy}, \citenamefont {El-Shewy}, \citenamefont {Abdo}, \citenamefont
  {Ould~Abdellahi},\ and\ \citenamefont {Al-Hagan}}]{EL-Shamy19}%
  \BibitemOpen
  \bibfield  {author} {\bibinfo {author} {\bibfnamefont {E.}~\bibnamefont
  {EL-Shamy}}, \bibinfo {author} {\bibfnamefont {E.}~\bibnamefont {El-Shewy}},
  \bibinfo {author} {\bibfnamefont {N.}~\bibnamefont {Abdo}}, \bibinfo {author}
  {\bibfnamefont {M.}~\bibnamefont {Ould~Abdellahi}}, \ and\ \bibinfo {author}
  {\bibfnamefont {O.}~\bibnamefont {Al-Hagan}},\ }\href {\doibase
  10.1002/ctpp.201800095} {\bibfield  {journal} {\bibinfo  {journal}
  {Contributions to Plasma Physics}\ }\textbf {\bibinfo {volume} {59}},\
  \bibinfo {pages} {304} (\bibinfo {year} {2019})}\BibitemShut {NoStop}%
\bibitem [{\citenamefont {Ur-Rehman}\ and\ \citenamefont
  {Mahmood}(2019)}]{Ur-Rehman19}%
  \BibitemOpen
  \bibfield  {author} {\bibinfo {author} {\bibfnamefont {H.}~\bibnamefont
  {Ur-Rehman}}\ and\ \bibinfo {author} {\bibfnamefont {S.}~\bibnamefont
  {Mahmood}},\ }\href {\doibase 10.1002/ctpp.201800037} {\bibfield  {journal}
  {\bibinfo  {journal} {Contributions to Plasma Physics}\ }\textbf {\bibinfo
  {volume} {59}},\ \bibinfo {pages} {236} (\bibinfo {year} {2019})}\BibitemShut
  {NoStop}%
\bibitem [{\citenamefont {Alam}\ and\ \citenamefont {Talukder}()}]{Alam19}%
  \BibitemOpen
  \bibfield  {author} {\bibinfo {author} {\bibfnamefont {M.~S.}\ \bibnamefont
  {Alam}}\ and\ \bibinfo {author} {\bibfnamefont {M.~R.}\ \bibnamefont
  {Talukder}},\ }\href {\doibase 10.1002/ctpp.201800163} {\bibfield  {journal}
  {\bibinfo  {journal} {Contributions to Plasma Physics}\ }\textbf {\bibinfo
  {volume} {59}},\ \bibinfo {pages} {e201800163}}\BibitemShut {NoStop}%
\bibitem [{\citenamefont {Wazwaz}(2015)}]{Wazwaz15}%
  \BibitemOpen
  \bibfield  {author} {\bibinfo {author} {\bibfnamefont {A.-M.}\ \bibnamefont
  {Wazwaz}},\ }\href {\doibase 10.1016/j.chaos.2015.03.018} {\bibfield
  {journal} {\bibinfo  {journal} {Chaos, Solitons \& Fractals}\ }\textbf
  {\bibinfo {volume} {76}},\ \bibinfo {pages} {93} (\bibinfo {year}
  {2015})}\BibitemShut {NoStop}%
\bibitem [{\citenamefont {Saini}\ \emph {et~al.}(2016)\citenamefont {Saini},
  \citenamefont {Kaur},\ and\ \citenamefont {Gill}}]{Saini16}%
  \BibitemOpen
  \bibfield  {author} {\bibinfo {author} {\bibfnamefont {N.~S.}\ \bibnamefont
  {Saini}}, \bibinfo {author} {\bibfnamefont {B.}~\bibnamefont {Kaur}}, \ and\
  \bibinfo {author} {\bibfnamefont {T.~S.}\ \bibnamefont {Gill}},\ }\href
  {\doibase 10.1063/1.4972542} {\bibfield  {journal} {\bibinfo  {journal}
  {Physics of Plasmas}\ }\textbf {\bibinfo {volume} {23}},\ \bibinfo {pages}
  {123705} (\bibinfo {year} {2016})}\BibitemShut {NoStop}%
\bibitem [{\citenamefont {Atteya}\ \emph {et~al.}(2018)\citenamefont {Atteya},
  \citenamefont {Sultana},\ and\ \citenamefont {Schlickeiser}}]{Atteya18}%
  \BibitemOpen
  \bibfield  {author} {\bibinfo {author} {\bibfnamefont {A.}~\bibnamefont
  {Atteya}}, \bibinfo {author} {\bibfnamefont {S.}~\bibnamefont {Sultana}}, \
  and\ \bibinfo {author} {\bibfnamefont {R.}~\bibnamefont {Schlickeiser}},\
  }\href {\doibase https://doi.org/10.1016/j.cjph.2018.09.002} {\bibfield
  {journal} {\bibinfo  {journal} {Chinese Journal of Physics}\ }\textbf
  {\bibinfo {volume} {56}},\ \bibinfo {pages} {1931 } (\bibinfo {year}
  {2018})}\BibitemShut {NoStop}%
\bibitem [{\citenamefont {Yahia}\ \emph {et~al.}(2019)\citenamefont {Yahia},
  \citenamefont {El-Labany}, \citenamefont {Sabry}, \citenamefont {Moslem},\
  and\ \citenamefont {Elghmaz}}]{Yahia19}%
  \BibitemOpen
  \bibfield  {author} {\bibinfo {author} {\bibfnamefont {M.~E.}\ \bibnamefont
  {Yahia}}, \bibinfo {author} {\bibfnamefont {S.~K.}\ \bibnamefont
  {El-Labany}}, \bibinfo {author} {\bibfnamefont {R.}~\bibnamefont {Sabry}},
  \bibinfo {author} {\bibfnamefont {W.~M.}\ \bibnamefont {Moslem}}, \ and\
  \bibinfo {author} {\bibfnamefont {E.~A.}\ \bibnamefont {Elghmaz}},\
  }\href@noop {} {\bibfield  {journal} {\bibinfo  {journal} {IEEE Transactions
  on Plasma Science}\ }\textbf {\bibinfo {volume} {47}},\ \bibinfo {pages}
  {762} (\bibinfo {year} {2019})}\BibitemShut {NoStop}%
\bibitem [{\citenamefont {Wang}\ \emph {et~al.}(2013)\citenamefont {Wang},
  \citenamefont {Li}, \citenamefont {Dai}, \citenamefont {Chen},\ and\
  \citenamefont {Zhang}}]{Wang13}%
  \BibitemOpen
  \bibfield  {author} {\bibinfo {author} {\bibfnamefont {Y.-Y.}\ \bibnamefont
  {Wang}}, \bibinfo {author} {\bibfnamefont {J.-T.}\ \bibnamefont {Li}},
  \bibinfo {author} {\bibfnamefont {C.-Q.}\ \bibnamefont {Dai}}, \bibinfo
  {author} {\bibfnamefont {X.-F.}\ \bibnamefont {Chen}}, \ and\ \bibinfo
  {author} {\bibfnamefont {J.-F.}\ \bibnamefont {Zhang}},\ }\href {\doibase
  10.1016/j.physleta.2013.06.008} {\bibfield  {journal} {\bibinfo  {journal}
  {Physics Letters A}\ }\textbf {\bibinfo {volume} {377}},\ \bibinfo {pages}
  {2097} (\bibinfo {year} {2013})}\BibitemShut {NoStop}%
\bibitem [{\citenamefont {Mukherjee}\ \emph {et~al.}(2015)\citenamefont
  {Mukherjee}, \citenamefont {Janaki},\ and\ \citenamefont
  {Kundu}}]{Mukherjee15}%
  \BibitemOpen
  \bibfield  {author} {\bibinfo {author} {\bibfnamefont {A.}~\bibnamefont
  {Mukherjee}}, \bibinfo {author} {\bibfnamefont {M.~S.}\ \bibnamefont
  {Janaki}}, \ and\ \bibinfo {author} {\bibfnamefont {A.}~\bibnamefont
  {Kundu}},\ }\href {\doibase 10.1063/1.4938513} {\bibfield  {journal}
  {\bibinfo  {journal} {Physics of Plasmas}\ }\textbf {\bibinfo {volume}
  {22}},\ \bibinfo {pages} {122114} (\bibinfo {year} {2015})}\BibitemShut
  {NoStop}%
\bibitem [{\citenamefont {Ishihara}\ \emph {et~al.}(2018)\citenamefont
  {Ishihara}, \citenamefont {Matsuno}, \citenamefont {Takahashi},\ and\
  \citenamefont {Teramae}}]{Ishihara18}%
  \BibitemOpen
  \bibfield  {author} {\bibinfo {author} {\bibfnamefont {H.}~\bibnamefont
  {Ishihara}}, \bibinfo {author} {\bibfnamefont {K.}~\bibnamefont {Matsuno}},
  \bibinfo {author} {\bibfnamefont {M.}~\bibnamefont {Takahashi}}, \ and\
  \bibinfo {author} {\bibfnamefont {S.}~\bibnamefont {Teramae}},\ }\href
  {\doibase 10.1103/PhysRevD.98.123010} {\bibfield  {journal} {\bibinfo
  {journal} {Phys. Rev. D}\ }\textbf {\bibinfo {volume} {98}},\ \bibinfo
  {pages} {123010} (\bibinfo {year} {2018})}\BibitemShut {NoStop}%
\bibitem [{\citenamefont {Samsonov}\ \emph {et~al.}(2002)\citenamefont
  {Samsonov}, \citenamefont {Ivlev}, \citenamefont {Quinn}, \citenamefont
  {Morfill},\ and\ \citenamefont {Zhdanov}}]{Samsonov02}%
  \BibitemOpen
  \bibfield  {author} {\bibinfo {author} {\bibfnamefont {D.}~\bibnamefont
  {Samsonov}}, \bibinfo {author} {\bibfnamefont {A.~V.}\ \bibnamefont {Ivlev}},
  \bibinfo {author} {\bibfnamefont {R.~A.}\ \bibnamefont {Quinn}}, \bibinfo
  {author} {\bibfnamefont {G.}~\bibnamefont {Morfill}}, \ and\ \bibinfo
  {author} {\bibfnamefont {S.}~\bibnamefont {Zhdanov}},\ }\href {\doibase
  10.1103/PhysRevLett.88.095004} {\bibfield  {journal} {\bibinfo  {journal}
  {Phys. Rev. Lett.}\ }\textbf {\bibinfo {volume} {88}},\ \bibinfo {pages}
  {095004} (\bibinfo {year} {2002})}\BibitemShut {NoStop}%
\bibitem [{\citenamefont {Nosenko}\ \emph {et~al.}(2004)\citenamefont
  {Nosenko}, \citenamefont {Avinash}, \citenamefont {Goree},\ and\
  \citenamefont {Liu}}]{Nosenko04}%
  \BibitemOpen
  \bibfield  {author} {\bibinfo {author} {\bibfnamefont {V.}~\bibnamefont
  {Nosenko}}, \bibinfo {author} {\bibfnamefont {K.}~\bibnamefont {Avinash}},
  \bibinfo {author} {\bibfnamefont {J.}~\bibnamefont {Goree}}, \ and\ \bibinfo
  {author} {\bibfnamefont {B.}~\bibnamefont {Liu}},\ }\href {\doibase
  10.1103/PhysRevLett.92.085001} {\bibfield  {journal} {\bibinfo  {journal}
  {Phys. Rev. Lett.}\ }\textbf {\bibinfo {volume} {92}},\ \bibinfo {pages}
  {085001} (\bibinfo {year} {2004})}\BibitemShut {NoStop}%
\bibitem [{\citenamefont {Nosenko}\ \emph {et~al.}(2006)\citenamefont
  {Nosenko}, \citenamefont {Goree},\ and\ \citenamefont {Skiff}}]{Nosenko06}%
  \BibitemOpen
  \bibfield  {author} {\bibinfo {author} {\bibfnamefont {V.}~\bibnamefont
  {Nosenko}}, \bibinfo {author} {\bibfnamefont {J.}~\bibnamefont {Goree}}, \
  and\ \bibinfo {author} {\bibfnamefont {F.}~\bibnamefont {Skiff}},\ }\href
  {\doibase 10.1103/PhysRevE.73.016401} {\bibfield  {journal} {\bibinfo
  {journal} {Phys. Rev. E}\ }\textbf {\bibinfo {volume} {73}},\ \bibinfo
  {pages} {016401} (\bibinfo {year} {2006})}\BibitemShut {NoStop}%
\bibitem [{\citenamefont {Sheridan}\ \emph {et~al.}(2008)\citenamefont
  {Sheridan}, \citenamefont {Nosenko},\ and\ \citenamefont
  {Goree}}]{Sheridan08}%
  \BibitemOpen
  \bibfield  {author} {\bibinfo {author} {\bibfnamefont {T.~E.}\ \bibnamefont
  {Sheridan}}, \bibinfo {author} {\bibfnamefont {V.}~\bibnamefont {Nosenko}}, \
  and\ \bibinfo {author} {\bibfnamefont {J.}~\bibnamefont {Goree}},\ }\href
  {\doibase 10.1063/1.2955476} {\bibfield  {journal} {\bibinfo  {journal}
  {Physics of Plasmas}\ }\textbf {\bibinfo {volume} {15}},\ \bibinfo {pages}
  {073703} (\bibinfo {year} {2008})}\BibitemShut {NoStop}%
\bibitem [{\citenamefont {Bandyopadhyay}\ \emph {et~al.}(2008)\citenamefont
  {Bandyopadhyay}, \citenamefont {Prasad}, \citenamefont {Sen},\ and\
  \citenamefont {Kaw}}]{Bandyopadhyay08}%
  \BibitemOpen
  \bibfield  {author} {\bibinfo {author} {\bibfnamefont {P.}~\bibnamefont
  {Bandyopadhyay}}, \bibinfo {author} {\bibfnamefont {G.}~\bibnamefont
  {Prasad}}, \bibinfo {author} {\bibfnamefont {A.}~\bibnamefont {Sen}}, \ and\
  \bibinfo {author} {\bibfnamefont {P.~K.}\ \bibnamefont {Kaw}},\ }\href
  {\doibase 10.1103/PhysRevLett.101.065006} {\bibfield  {journal} {\bibinfo
  {journal} {Phys. Rev. Lett.}\ }\textbf {\bibinfo {volume} {101}},\ \bibinfo
  {pages} {065006} (\bibinfo {year} {2008})}\BibitemShut {NoStop}%
\bibitem [{\citenamefont {Heidemann}\ \emph {et~al.}(2009)\citenamefont
  {Heidemann}, \citenamefont {Zhdanov}, \citenamefont {S\"utterlin},
  \citenamefont {Thomas},\ and\ \citenamefont {Morfill}}]{Heidemann09}%
  \BibitemOpen
  \bibfield  {author} {\bibinfo {author} {\bibfnamefont {R.}~\bibnamefont
  {Heidemann}}, \bibinfo {author} {\bibfnamefont {S.}~\bibnamefont {Zhdanov}},
  \bibinfo {author} {\bibfnamefont {R.}~\bibnamefont {S\"utterlin}}, \bibinfo
  {author} {\bibfnamefont {H.~M.}\ \bibnamefont {Thomas}}, \ and\ \bibinfo
  {author} {\bibfnamefont {G.~E.}\ \bibnamefont {Morfill}},\ }\href {\doibase
  10.1103/PhysRevLett.102.135002} {\bibfield  {journal} {\bibinfo  {journal}
  {Phys. Rev. Lett.}\ }\textbf {\bibinfo {volume} {102}},\ \bibinfo {pages}
  {135002} (\bibinfo {year} {2009})}\BibitemShut {NoStop}%
\bibitem [{\citenamefont {Zhdanov}\ \emph {et~al.}(2010)\citenamefont
  {Zhdanov}, \citenamefont {Heidemann}, \citenamefont {Thoma}, \citenamefont
  {S\"utterlin}, \citenamefont {Thomas}, \citenamefont {H\"ofner},
  \citenamefont {Tarantik}, \citenamefont {Morfill}, \citenamefont {Usachev},
  \citenamefont {Petrov},\ and\ \citenamefont {Fortov}}]{Zhdanov10}%
  \BibitemOpen
  \bibfield  {author} {\bibinfo {author} {\bibfnamefont {S.}~\bibnamefont
  {Zhdanov}}, \bibinfo {author} {\bibfnamefont {R.}~\bibnamefont {Heidemann}},
  \bibinfo {author} {\bibfnamefont {M.~H.}\ \bibnamefont {Thoma}}, \bibinfo
  {author} {\bibfnamefont {R.}~\bibnamefont {S\"utterlin}}, \bibinfo {author}
  {\bibfnamefont {H.~M.}\ \bibnamefont {Thomas}}, \bibinfo {author}
  {\bibfnamefont {H.}~\bibnamefont {H\"ofner}}, \bibinfo {author}
  {\bibfnamefont {K.}~\bibnamefont {Tarantik}}, \bibinfo {author}
  {\bibfnamefont {G.~E.}\ \bibnamefont {Morfill}}, \bibinfo {author}
  {\bibfnamefont {A.~D.}\ \bibnamefont {Usachev}}, \bibinfo {author}
  {\bibfnamefont {O.~F.}\ \bibnamefont {Petrov}}, \ and\ \bibinfo {author}
  {\bibfnamefont {V.~E.}\ \bibnamefont {Fortov}},\ }\href {\doibase
  10.1209/0295-5075/89/25001} {\bibfield  {journal} {\bibinfo  {journal} {{EPL}
  (Europhysics Letters)}\ }\textbf {\bibinfo {volume} {89}},\ \bibinfo {pages}
  {25001} (\bibinfo {year} {2010})}\BibitemShut {NoStop}%
\bibitem [{\citenamefont {Boruah}\ \emph {et~al.}(2015)\citenamefont {Boruah},
  \citenamefont {Sharma}, \citenamefont {Bailung},\ and\ \citenamefont
  {Nakamura}}]{Boruah15}%
  \BibitemOpen
  \bibfield  {author} {\bibinfo {author} {\bibfnamefont {A.}~\bibnamefont
  {Boruah}}, \bibinfo {author} {\bibfnamefont {S.~K.}\ \bibnamefont {Sharma}},
  \bibinfo {author} {\bibfnamefont {H.}~\bibnamefont {Bailung}}, \ and\
  \bibinfo {author} {\bibfnamefont {Y.}~\bibnamefont {Nakamura}},\ }\href
  {\doibase 10.1063/1.4931735} {\bibfield  {journal} {\bibinfo  {journal}
  {Physics of Plasmas}\ }\textbf {\bibinfo {volume} {22}},\ \bibinfo {pages}
  {093706} (\bibinfo {year} {2015})}\BibitemShut {NoStop}%
\bibitem [{\citenamefont {Arora}\ \emph {et~al.}(2019)\citenamefont {Arora},
  \citenamefont {Bandyopadhyay}, \citenamefont {Hariprasad},\ and\
  \citenamefont {Sen}}]{Arora19}%
  \BibitemOpen
  \bibfield  {author} {\bibinfo {author} {\bibfnamefont {G.}~\bibnamefont
  {Arora}}, \bibinfo {author} {\bibfnamefont {P.}~\bibnamefont
  {Bandyopadhyay}}, \bibinfo {author} {\bibfnamefont {M.~G.}\ \bibnamefont
  {Hariprasad}}, \ and\ \bibinfo {author} {\bibfnamefont {A.}~\bibnamefont
  {Sen}},\ }\href {\doibase 10.1063/1.5115313} {\bibfield  {journal} {\bibinfo
  {journal} {Physics of Plasmas}\ }\textbf {\bibinfo {volume} {26}},\ \bibinfo
  {pages} {093701} (\bibinfo {year} {2019})}\BibitemShut {NoStop}%
\bibitem [{\citenamefont {Usachev}\ \emph {et~al.}(2014)\citenamefont
  {Usachev}, \citenamefont {Zobnin}, \citenamefont {Petrov}, \citenamefont
  {Fortov}, \citenamefont {Thoma}, \citenamefont {Höfner}, \citenamefont
  {Fink}, \citenamefont {Ivlev},\ and\ \citenamefont {Morfill}}]{Usachev14}%
  \BibitemOpen
  \bibfield  {author} {\bibinfo {author} {\bibfnamefont {A.}~\bibnamefont
  {Usachev}}, \bibinfo {author} {\bibfnamefont {A.}~\bibnamefont {Zobnin}},
  \bibinfo {author} {\bibfnamefont {O.}~\bibnamefont {Petrov}}, \bibinfo
  {author} {\bibfnamefont {V.}~\bibnamefont {Fortov}}, \bibinfo {author}
  {\bibfnamefont {M.~H.}\ \bibnamefont {Thoma}}, \bibinfo {author}
  {\bibfnamefont {H.}~\bibnamefont {Höfner}}, \bibinfo {author} {\bibfnamefont
  {M.}~\bibnamefont {Fink}}, \bibinfo {author} {\bibfnamefont {A.}~\bibnamefont
  {Ivlev}}, \ and\ \bibinfo {author} {\bibfnamefont {G.}~\bibnamefont
  {Morfill}},\ }\href {\doibase 10.1088/1367-2630/16/5/053028} {\bibfield
  {journal} {\bibinfo  {journal} {New Journal of Physics}\ }\textbf {\bibinfo
  {volume} {16}},\ \bibinfo {pages} {053028} (\bibinfo {year}
  {2014})}\BibitemShut {NoStop}%
\bibitem [{\citenamefont {Sheridan}\ and\ \citenamefont
  {Gallagher}(2017)}]{Sheridan17}%
  \BibitemOpen
  \bibfield  {author} {\bibinfo {author} {\bibfnamefont {T.~E.}\ \bibnamefont
  {Sheridan}}\ and\ \bibinfo {author} {\bibfnamefont {J.~C.}\ \bibnamefont
  {Gallagher}},\ }\href {\doibase 10.1017/S0022377817000411} {\bibfield
  {journal} {\bibinfo  {journal} {Journal of Plasma Physics}\ }\textbf
  {\bibinfo {volume} {83}},\ \bibinfo {pages} {905830305} (\bibinfo {year}
  {2017})}\BibitemShut {NoStop}%
\bibitem [{\citenamefont {Berbri}\ and\ \citenamefont
  {Tribeche}(2009)}]{Berbri09}%
  \BibitemOpen
  \bibfield  {author} {\bibinfo {author} {\bibfnamefont {A.}~\bibnamefont
  {Berbri}}\ and\ \bibinfo {author} {\bibfnamefont {M.}~\bibnamefont
  {Tribeche}},\ }\href {\doibase 10.1063/1.3133184} {\bibfield  {journal}
  {\bibinfo  {journal} {Physics of Plasmas}\ }\textbf {\bibinfo {volume}
  {16}},\ \bibinfo {pages} {053703} (\bibinfo {year} {2009})}\BibitemShut
  {NoStop}%
\bibitem [{\citenamefont {Nouri~Kadijani}\ and\ \citenamefont
  {Zaremoghaddam}(2012)}]{Nouri12}%
  \BibitemOpen
  \bibfield  {author} {\bibinfo {author} {\bibfnamefont {M.}~\bibnamefont
  {Nouri~Kadijani}}\ and\ \bibinfo {author} {\bibfnamefont {H.}~\bibnamefont
  {Zaremoghaddam}},\ }\href {\doibase 10.1007/s10894-011-9492-2} {\bibfield
  {journal} {\bibinfo  {journal} {Journal of Fusion Energy}\ }\textbf {\bibinfo
  {volume} {31}},\ \bibinfo {pages} {455} (\bibinfo {year} {2012})}\BibitemShut
  {NoStop}%
\bibitem [{\citenamefont {Chatterjee}\ \emph {et~al.}(2018)\citenamefont
  {Chatterjee}, \citenamefont {Ali},\ and\ \citenamefont
  {Saha}}]{Chatterjee18}%
  \BibitemOpen
  \bibfield  {author} {\bibinfo {author} {\bibfnamefont {P.}~\bibnamefont
  {Chatterjee}}, \bibinfo {author} {\bibfnamefont {R.}~\bibnamefont {Ali}}, \
  and\ \bibinfo {author} {\bibfnamefont {A.}~\bibnamefont {Saha}},\ }\href
  {\doibase 10.1515/zna-2017-0358} {\bibfield  {journal} {\bibinfo  {journal}
  {Zeitschrift für Naturforschung A}\ }\textbf {\bibinfo {volume} {73}},\
  \bibinfo {pages} {151} (\bibinfo {year} {2018})}\BibitemShut {NoStop}%
\bibitem [{\citenamefont {Rahman}\ \emph {et~al.}(2018)\citenamefont {Rahman},
  \citenamefont {Chowdhury}, \citenamefont {Mannan}, \citenamefont {Rahman},\
  and\ \citenamefont {Mamun}}]{Rahman18}%
  \BibitemOpen
  \bibfield  {author} {\bibinfo {author} {\bibfnamefont {M.}~\bibnamefont
  {Rahman}}, \bibinfo {author} {\bibfnamefont {N.}~\bibnamefont {Chowdhury}},
  \bibinfo {author} {\bibfnamefont {A.}~\bibnamefont {Mannan}}, \bibinfo
  {author} {\bibfnamefont {M.}~\bibnamefont {Rahman}}, \ and\ \bibinfo {author}
  {\bibfnamefont {A.}~\bibnamefont {Mamun}},\ }\href {\doibase
  10.1016/j.cjph.2018.09.020} {\bibfield  {journal} {\bibinfo  {journal}
  {Chinese Journal of Physics}\ }\textbf {\bibinfo {volume} {56}},\ \bibinfo
  {pages} {2061} (\bibinfo {year} {2018})}\BibitemShut {NoStop}%
\bibitem [{\citenamefont {Paul}\ \emph {et~al.}(2019)\citenamefont {Paul},
  \citenamefont {Mondal},\ and\ \citenamefont {Chatterjee}}]{Paul19}%
  \BibitemOpen
  \bibfield  {author} {\bibinfo {author} {\bibfnamefont {N.}~\bibnamefont
  {Paul}}, \bibinfo {author} {\bibfnamefont {K.}~\bibnamefont {Mondal}}, \ and\
  \bibinfo {author} {\bibfnamefont {P.}~\bibnamefont {Chatterjee}},\ }\href
  {\doibase 10.1515/zna-2018-0519} {\bibfield  {journal} {\bibinfo  {journal}
  {Z. Naturforsch. A}\ }\textbf {\bibinfo {volume} {74}},\ \bibinfo {pages}
  {861} (\bibinfo {year} {2019})}\BibitemShut {NoStop}%
\bibitem [{\citenamefont {Gao}\ \emph {et~al.}(2019)\citenamefont {Gao},
  \citenamefont {Zhang}, \citenamefont {Wu}, \citenamefont {Luo}, \citenamefont
  {Duan},\ and\ \citenamefont {Li}}]{Gao19}%
  \BibitemOpen
  \bibfield  {author} {\bibinfo {author} {\bibfnamefont {D.-N.}\ \bibnamefont
  {Gao}}, \bibinfo {author} {\bibfnamefont {Z.-R.}\ \bibnamefont {Zhang}},
  \bibinfo {author} {\bibfnamefont {J.-P.}\ \bibnamefont {Wu}}, \bibinfo
  {author} {\bibfnamefont {D.}~\bibnamefont {Luo}}, \bibinfo {author}
  {\bibfnamefont {W.-S.}\ \bibnamefont {Duan}}, \ and\ \bibinfo {author}
  {\bibfnamefont {Z.-Z.}\ \bibnamefont {Li}},\ }\href {\doibase
  10.1007/s13538-019-00687-0} {\bibfield  {journal} {\bibinfo  {journal}
  {Brazilian Journal of Physics}\ }\textbf {\bibinfo {volume} {49}},\ \bibinfo
  {pages} {693} (\bibinfo {year} {2019})}\BibitemShut {NoStop}%
\bibitem [{\citenamefont {Dzhumagulova}\ \emph {et~al.}(2012)\citenamefont
  {Dzhumagulova}, \citenamefont {Ramazanov},\ and\ \citenamefont
  {Masheeva}}]{Langevin}%
  \BibitemOpen
  \bibfield  {author} {\bibinfo {author} {\bibfnamefont {K.~N.}\ \bibnamefont
  {Dzhumagulova}}, \bibinfo {author} {\bibfnamefont {T.~S.}\ \bibnamefont
  {Ramazanov}}, \ and\ \bibinfo {author} {\bibfnamefont {R.~U.}\ \bibnamefont
  {Masheeva}},\ }\href {\doibase 10.1002/ctpp.201100070} {\bibfield  {journal}
  {\bibinfo  {journal} {Contributions to Plasma Physics}\ }\textbf {\bibinfo
  {volume} {52}},\ \bibinfo {pages} {182} (\bibinfo {year} {2012})}\BibitemShut
  {NoStop}%
\bibitem [{\citenamefont {Avinash}\ \emph {et~al.}(2003)\citenamefont
  {Avinash}, \citenamefont {Zhu}, \citenamefont {Nosenko},\ and\ \citenamefont
  {Goree}}]{Avinash03}%
  \BibitemOpen
  \bibfield  {author} {\bibinfo {author} {\bibfnamefont {K.}~\bibnamefont
  {Avinash}}, \bibinfo {author} {\bibfnamefont {P.}~\bibnamefont {Zhu}},
  \bibinfo {author} {\bibfnamefont {V.}~\bibnamefont {Nosenko}}, \ and\
  \bibinfo {author} {\bibfnamefont {J.}~\bibnamefont {Goree}},\ }\href
  {\doibase 10.1103/PhysRevE.68.046402} {\bibfield  {journal} {\bibinfo
  {journal} {Phys. Rev. E}\ }\textbf {\bibinfo {volume} {68}},\ \bibinfo
  {pages} {046402} (\bibinfo {year} {2003})}\BibitemShut {NoStop}%
\bibitem [{\citenamefont {Kumar}\ \emph {et~al.}(2017)\citenamefont {Kumar},
  \citenamefont {Tiwari},\ and\ \citenamefont {Das}}]{Kumar17}%
  \BibitemOpen
  \bibfield  {author} {\bibinfo {author} {\bibfnamefont {S.}~\bibnamefont
  {Kumar}}, \bibinfo {author} {\bibfnamefont {S.~K.}\ \bibnamefont {Tiwari}}, \
  and\ \bibinfo {author} {\bibfnamefont {A.}~\bibnamefont {Das}},\ }\href
  {\doibase 10.1063/1.4978779} {\bibfield  {journal} {\bibinfo  {journal}
  {Physics of Plasmas}\ }\textbf {\bibinfo {volume} {24}},\ \bibinfo {pages}
  {033711} (\bibinfo {year} {2017})}\BibitemShut {NoStop}%
\bibitem [{\citenamefont {Tiwari}\ and\ \citenamefont {Sen}(2016)}]{Tiwari16}%
  \BibitemOpen
  \bibfield  {author} {\bibinfo {author} {\bibfnamefont {S.~K.}\ \bibnamefont
  {Tiwari}}\ and\ \bibinfo {author} {\bibfnamefont {A.}~\bibnamefont {Sen}},\
  }\href {\doibase 10.1063/1.4964908} {\bibfield  {journal} {\bibinfo
  {journal} {Physics of Plasmas}\ }\textbf {\bibinfo {volume} {23}},\ \bibinfo
  {pages} {100705} (\bibinfo {year} {2016})}\BibitemShut {NoStop}%
\bibitem [{\citenamefont {Tiwari}\ \emph {et~al.}(2015)\citenamefont {Tiwari},
  \citenamefont {Das}, \citenamefont {Sen},\ and\ \citenamefont
  {Kaw}}]{Tiwari15}%
  \BibitemOpen
  \bibfield  {author} {\bibinfo {author} {\bibfnamefont {S.~K.}\ \bibnamefont
  {Tiwari}}, \bibinfo {author} {\bibfnamefont {A.}~\bibnamefont {Das}},
  \bibinfo {author} {\bibfnamefont {A.}~\bibnamefont {Sen}}, \ and\ \bibinfo
  {author} {\bibfnamefont {P.}~\bibnamefont {Kaw}},\ }\href {\doibase
  10.1063/1.4916576} {\bibfield  {journal} {\bibinfo  {journal} {Physics of
  Plasmas}\ }\textbf {\bibinfo {volume} {22}},\ \bibinfo {pages} {033706}
  (\bibinfo {year} {2015})}\BibitemShut {NoStop}%
\bibitem [{\citenamefont {Chu}\ and\ \citenamefont {I}(1994)}]{Chu94}%
  \BibitemOpen
  \bibfield  {author} {\bibinfo {author} {\bibfnamefont {J.~H.}\ \bibnamefont
  {Chu}}\ and\ \bibinfo {author} {\bibfnamefont {L.}~\bibnamefont {I}},\ }\href
  {\doibase 10.1103/PhysRevLett.72.4009} {\bibfield  {journal} {\bibinfo
  {journal} {Phys. Rev. Lett.}\ }\textbf {\bibinfo {volume} {72}},\ \bibinfo
  {pages} {4009} (\bibinfo {year} {1994})}\BibitemShut {NoStop}%
\bibitem [{\citenamefont {Thomas}\ \emph {et~al.}(1994)\citenamefont {Thomas},
  \citenamefont {Morfill}, \citenamefont {Demmel}, \citenamefont {Goree},
  \citenamefont {Feuerbacher},\ and\ \citenamefont {M\"ohlmann}}]{Thomas94}%
  \BibitemOpen
  \bibfield  {author} {\bibinfo {author} {\bibfnamefont {H.}~\bibnamefont
  {Thomas}}, \bibinfo {author} {\bibfnamefont {G.~E.}\ \bibnamefont {Morfill}},
  \bibinfo {author} {\bibfnamefont {V.}~\bibnamefont {Demmel}}, \bibinfo
  {author} {\bibfnamefont {J.}~\bibnamefont {Goree}}, \bibinfo {author}
  {\bibfnamefont {B.}~\bibnamefont {Feuerbacher}}, \ and\ \bibinfo {author}
  {\bibfnamefont {D.}~\bibnamefont {M\"ohlmann}},\ }\href {\doibase
  10.1103/PhysRevLett.73.652} {\bibfield  {journal} {\bibinfo  {journal} {Phys.
  Rev. Lett.}\ }\textbf {\bibinfo {volume} {73}},\ \bibinfo {pages} {652}
  (\bibinfo {year} {1994})}\BibitemShut {NoStop}%
\bibitem [{\citenamefont {Melzer}\ \emph {et~al.}(1994)\citenamefont {Melzer},
  \citenamefont {Trottenberg},\ and\ \citenamefont {Piel}}]{Melzer94}%
  \BibitemOpen
  \bibfield  {author} {\bibinfo {author} {\bibfnamefont {A.}~\bibnamefont
  {Melzer}}, \bibinfo {author} {\bibfnamefont {T.}~\bibnamefont {Trottenberg}},
  \ and\ \bibinfo {author} {\bibfnamefont {A.}~\bibnamefont {Piel}},\ }\href
  {\doibase https://doi.org/10.1016/0375-9601(94)90144-9} {\bibfield  {journal}
  {\bibinfo  {journal} {Physics Letters A}\ }\textbf {\bibinfo {volume}
  {191}},\ \bibinfo {pages} {301} (\bibinfo {year} {1994})}\BibitemShut
  {NoStop}%
\bibitem [{\citenamefont {Baalrud}\ and\ \citenamefont
  {Daligault}(2017)}]{Baalrud17}%
  \BibitemOpen
  \bibfield  {author} {\bibinfo {author} {\bibfnamefont {S.~D.}\ \bibnamefont
  {Baalrud}}\ and\ \bibinfo {author} {\bibfnamefont {J.}~\bibnamefont
  {Daligault}},\ }\href {\doibase 10.1103/PhysRevE.96.043202} {\bibfield
  {journal} {\bibinfo  {journal} {Phys. Rev. E}\ }\textbf {\bibinfo {volume}
  {96}},\ \bibinfo {pages} {043202} (\bibinfo {year} {2017})}\BibitemShut
  {NoStop}%
\bibitem [{\citenamefont {Feng}\ \emph {et~al.}(2017)\citenamefont {Feng},
  \citenamefont {Lin},\ and\ \citenamefont {Murillo}}]{Feng17}%
  \BibitemOpen
  \bibfield  {author} {\bibinfo {author} {\bibfnamefont {Y.}~\bibnamefont
  {Feng}}, \bibinfo {author} {\bibfnamefont {W.}~\bibnamefont {Lin}}, \ and\
  \bibinfo {author} {\bibfnamefont {M.~S.}\ \bibnamefont {Murillo}},\ }\href
  {\doibase 10.1103/PhysRevE.96.053208} {\bibfield  {journal} {\bibinfo
  {journal} {Phys. Rev. E}\ }\textbf {\bibinfo {volume} {96}},\ \bibinfo
  {pages} {053208} (\bibinfo {year} {2017})}\BibitemShut {NoStop}%
\bibitem [{\citenamefont {Hartmann}\ \emph {et~al.}(2019)\citenamefont
  {Hartmann}, \citenamefont {Reyes}, \citenamefont {Kostadinova}, \citenamefont
  {Matthews}, \citenamefont {Hyde}, \citenamefont {Masheyeva}, \citenamefont
  {Dzhumagulova}, \citenamefont {Ramazanov}, \citenamefont {Ott}, \citenamefont
  {K\"ahlert}, \citenamefont {Bonitz}, \citenamefont {Korolov},\ and\
  \citenamefont {Donk\'o}}]{Hartmann19}%
  \BibitemOpen
  \bibfield  {author} {\bibinfo {author} {\bibfnamefont {P.}~\bibnamefont
  {Hartmann}}, \bibinfo {author} {\bibfnamefont {J.~C.}\ \bibnamefont {Reyes}},
  \bibinfo {author} {\bibfnamefont {E.~G.}\ \bibnamefont {Kostadinova}},
  \bibinfo {author} {\bibfnamefont {L.~S.}\ \bibnamefont {Matthews}}, \bibinfo
  {author} {\bibfnamefont {T.~W.}\ \bibnamefont {Hyde}}, \bibinfo {author}
  {\bibfnamefont {R.~U.}\ \bibnamefont {Masheyeva}}, \bibinfo {author}
  {\bibfnamefont {K.~N.}\ \bibnamefont {Dzhumagulova}}, \bibinfo {author}
  {\bibfnamefont {T.~S.}\ \bibnamefont {Ramazanov}}, \bibinfo {author}
  {\bibfnamefont {T.}~\bibnamefont {Ott}}, \bibinfo {author} {\bibfnamefont
  {H.}~\bibnamefont {K\"ahlert}}, \bibinfo {author} {\bibfnamefont
  {M.}~\bibnamefont {Bonitz}}, \bibinfo {author} {\bibfnamefont
  {I.}~\bibnamefont {Korolov}}, \ and\ \bibinfo {author} {\bibfnamefont
  {Z.}~\bibnamefont {Donk\'o}},\ }\href {\doibase 10.1103/PhysRevE.99.013203}
  {\bibfield  {journal} {\bibinfo  {journal} {Phys. Rev. E}\ }\textbf {\bibinfo
  {volume} {99}},\ \bibinfo {pages} {013203} (\bibinfo {year}
  {2019})}\BibitemShut {NoStop}%
\bibitem [{\citenamefont {{Scheiner}}\ and\ \citenamefont
  {{Baalrud}}(2019)}]{Scheiner19}%
  \BibitemOpen
  \bibfield  {author} {\bibinfo {author} {\bibfnamefont {B.}~\bibnamefont
  {{Scheiner}}}\ and\ \bibinfo {author} {\bibfnamefont {S.~D.}\ \bibnamefont
  {{Baalrud}}},\ }\href@noop {} {\bibfield  {journal} {\bibinfo  {journal}
  {arXiv e-prints}\ ,\ \bibinfo {eid} {arXiv:1908.08415}} (\bibinfo {year}
  {2019})}\BibitemShut {NoStop}%
\bibitem [{\citenamefont {Meyer}\ \emph {et~al.}(2017)\citenamefont {Meyer},
  \citenamefont {Laut}, \citenamefont {Zhdanov}, \citenamefont {Nosenko},\ and\
  \citenamefont {Thomas}}]{Meyer17}%
  \BibitemOpen
  \bibfield  {author} {\bibinfo {author} {\bibfnamefont {J.~K.}\ \bibnamefont
  {Meyer}}, \bibinfo {author} {\bibfnamefont {I.}~\bibnamefont {Laut}},
  \bibinfo {author} {\bibfnamefont {S.~K.}\ \bibnamefont {Zhdanov}}, \bibinfo
  {author} {\bibfnamefont {V.}~\bibnamefont {Nosenko}}, \ and\ \bibinfo
  {author} {\bibfnamefont {H.~M.}\ \bibnamefont {Thomas}},\ }\href {\doibase
  10.1103/PhysRevLett.119.255001} {\bibfield  {journal} {\bibinfo  {journal}
  {Phys. Rev. Lett.}\ }\textbf {\bibinfo {volume} {119}},\ \bibinfo {pages}
  {255001} (\bibinfo {year} {2017})}\BibitemShut {NoStop}%
\bibitem [{\citenamefont {Kumar}\ and\ \citenamefont {Das}(2018)}]{Kumar18}%
  \BibitemOpen
  \bibfield  {author} {\bibinfo {author} {\bibfnamefont {S.}~\bibnamefont
  {Kumar}}\ and\ \bibinfo {author} {\bibfnamefont {A.}~\bibnamefont {Das}},\
  }\href {\doibase 10.1103/PhysRevE.97.063202} {\bibfield  {journal} {\bibinfo
  {journal} {Phys. Rev. E}\ }\textbf {\bibinfo {volume} {97}},\ \bibinfo
  {pages} {063202} (\bibinfo {year} {2018})}\BibitemShut {NoStop}%
\bibitem [{\citenamefont {Li}\ \emph {et~al.}(2018)\citenamefont {Li},
  \citenamefont {Huang}, \citenamefont {Wang}, \citenamefont {Reichhardt},
  \citenamefont {Reichhardt}, \citenamefont {Murillo},\ and\ \citenamefont
  {Feng}}]{Li18}%
  \BibitemOpen
  \bibfield  {author} {\bibinfo {author} {\bibfnamefont {W.}~\bibnamefont
  {Li}}, \bibinfo {author} {\bibfnamefont {D.}~\bibnamefont {Huang}}, \bibinfo
  {author} {\bibfnamefont {K.}~\bibnamefont {Wang}}, \bibinfo {author}
  {\bibfnamefont {C.}~\bibnamefont {Reichhardt}}, \bibinfo {author}
  {\bibfnamefont {C.~J.~O.}\ \bibnamefont {Reichhardt}}, \bibinfo {author}
  {\bibfnamefont {M.~S.}\ \bibnamefont {Murillo}}, \ and\ \bibinfo {author}
  {\bibfnamefont {Y.}~\bibnamefont {Feng}},\ }\href {\doibase
  10.1103/PhysRevE.98.063203} {\bibfield  {journal} {\bibinfo  {journal} {Phys.
  Rev. E}\ }\textbf {\bibinfo {volume} {98}},\ \bibinfo {pages} {063203}
  (\bibinfo {year} {2018})}\BibitemShut {NoStop}%
\bibitem [{\citenamefont {Wong}\ \emph {et~al.}(2018)\citenamefont {Wong},
  \citenamefont {Goree},\ and\ \citenamefont {Haralson}}]{Wong18}%
  \BibitemOpen
  \bibfield  {author} {\bibinfo {author} {\bibfnamefont {C.-S.}\ \bibnamefont
  {Wong}}, \bibinfo {author} {\bibfnamefont {J.}~\bibnamefont {Goree}}, \ and\
  \bibinfo {author} {\bibfnamefont {Z.}~\bibnamefont {Haralson}},\ }\href
  {\doibase 10.1103/PhysRevE.98.063201} {\bibfield  {journal} {\bibinfo
  {journal} {Phys. Rev. E}\ }\textbf {\bibinfo {volume} {98}},\ \bibinfo
  {pages} {063201} (\bibinfo {year} {2018})}\BibitemShut {NoStop}%
\bibitem [{\citenamefont {Kryuchkov}\ \emph {et~al.}(2018)\citenamefont
  {Kryuchkov}, \citenamefont {Yakovlev}, \citenamefont {Gorbunov},
  \citenamefont {Cou\"edel}, \citenamefont {Lipaev},\ and\ \citenamefont
  {Yurchenko}}]{Kryuchkov18}%
  \BibitemOpen
  \bibfield  {author} {\bibinfo {author} {\bibfnamefont {N.~P.}\ \bibnamefont
  {Kryuchkov}}, \bibinfo {author} {\bibfnamefont {E.~V.}\ \bibnamefont
  {Yakovlev}}, \bibinfo {author} {\bibfnamefont {E.~A.}\ \bibnamefont
  {Gorbunov}}, \bibinfo {author} {\bibfnamefont {L.}~\bibnamefont {Cou\"edel}},
  \bibinfo {author} {\bibfnamefont {A.~M.}\ \bibnamefont {Lipaev}}, \ and\
  \bibinfo {author} {\bibfnamefont {S.~O.}\ \bibnamefont {Yurchenko}},\ }\href
  {\doibase 10.1103/PhysRevLett.121.075003} {\bibfield  {journal} {\bibinfo
  {journal} {Phys. Rev. Lett.}\ }\textbf {\bibinfo {volume} {121}},\ \bibinfo
  {pages} {075003} (\bibinfo {year} {2018})}\BibitemShut {NoStop}%
\bibitem [{\citenamefont {Hu}\ \emph {et~al.}(2019)\citenamefont {Hu},
  \citenamefont {Wang},\ and\ \citenamefont {I}}]{Hu19}%
  \BibitemOpen
  \bibfield  {author} {\bibinfo {author} {\bibfnamefont {H.-W.}\ \bibnamefont
  {Hu}}, \bibinfo {author} {\bibfnamefont {W.}~\bibnamefont {Wang}}, \ and\
  \bibinfo {author} {\bibfnamefont {L.}~\bibnamefont {I}},\ }\href {\doibase
  10.1103/PhysRevLett.123.065002} {\bibfield  {journal} {\bibinfo  {journal}
  {Phys. Rev. Lett.}\ }\textbf {\bibinfo {volume} {123}},\ \bibinfo {pages}
  {065002} (\bibinfo {year} {2019})}\BibitemShut {NoStop}%
\bibitem [{\citenamefont {Hartmann}\ \emph {et~al.}(2010)\citenamefont
  {Hartmann}, \citenamefont {Douglass}, \citenamefont {Reyes}, \citenamefont
  {Matthews}, \citenamefont {Hyde}, \citenamefont {Kov\'acs},\ and\
  \citenamefont {Donk\'o}}]{Hartmann10}%
  \BibitemOpen
  \bibfield  {author} {\bibinfo {author} {\bibfnamefont {P.}~\bibnamefont
  {Hartmann}}, \bibinfo {author} {\bibfnamefont {A.}~\bibnamefont {Douglass}},
  \bibinfo {author} {\bibfnamefont {J.~C.}\ \bibnamefont {Reyes}}, \bibinfo
  {author} {\bibfnamefont {L.~S.}\ \bibnamefont {Matthews}}, \bibinfo {author}
  {\bibfnamefont {T.~W.}\ \bibnamefont {Hyde}}, \bibinfo {author}
  {\bibfnamefont {A.}~\bibnamefont {Kov\'acs}}, \ and\ \bibinfo {author}
  {\bibfnamefont {Z.}~\bibnamefont {Donk\'o}},\ }\href {\doibase
  10.1103/PhysRevLett.105.115004} {\bibfield  {journal} {\bibinfo  {journal}
  {Phys. Rev. Lett.}\ }\textbf {\bibinfo {volume} {105}},\ \bibinfo {pages}
  {115004} (\bibinfo {year} {2010})}\BibitemShut {NoStop}%
\bibitem [{\citenamefont {Su}\ \emph {et~al.}(2012)\citenamefont {Su},
  \citenamefont {Io},\ and\ \citenamefont {I}}]{Su12}%
  \BibitemOpen
  \bibfield  {author} {\bibinfo {author} {\bibfnamefont {Y.-S.}\ \bibnamefont
  {Su}}, \bibinfo {author} {\bibfnamefont {C.-W.}\ \bibnamefont {Io}}, \ and\
  \bibinfo {author} {\bibfnamefont {L.}~\bibnamefont {I}},\ }\href {\doibase
  10.1103/PhysRevE.86.016405} {\bibfield  {journal} {\bibinfo  {journal} {Phys.
  Rev. E}\ }\textbf {\bibinfo {volume} {86}},\ \bibinfo {pages} {016405}
  (\bibinfo {year} {2012})}\BibitemShut {NoStop}%
\bibitem [{\citenamefont {Petrov}\ \emph {et~al.}(2015)\citenamefont {Petrov},
  \citenamefont {Vasiliev}, \citenamefont {Tun}, \citenamefont {Statsenko},
  \citenamefont {Vaulina}, \citenamefont {Vasilieva},\ and\ \citenamefont
  {Fortov}}]{Petrov2015}%
  \BibitemOpen
  \bibfield  {author} {\bibinfo {author} {\bibfnamefont {O.~F.}\ \bibnamefont
  {Petrov}}, \bibinfo {author} {\bibfnamefont {M.~M.}\ \bibnamefont
  {Vasiliev}}, \bibinfo {author} {\bibfnamefont {Y.}~\bibnamefont {Tun}},
  \bibinfo {author} {\bibfnamefont {K.~B.}\ \bibnamefont {Statsenko}}, \bibinfo
  {author} {\bibfnamefont {O.~S.}\ \bibnamefont {Vaulina}}, \bibinfo {author}
  {\bibfnamefont {E.~V.}\ \bibnamefont {Vasilieva}}, \ and\ \bibinfo {author}
  {\bibfnamefont {V.~E.}\ \bibnamefont {Fortov}},\ }\href {\doibase
  10.1134/S1063776115020181} {\bibfield  {journal} {\bibinfo  {journal}
  {Journal of Experimental and Theoretical Physics}\ }\textbf {\bibinfo
  {volume} {120}},\ \bibinfo {pages} {327} (\bibinfo {year}
  {2015})}\BibitemShut {NoStop}%
\bibitem [{\citenamefont {Jaiswal}\ and\ \citenamefont
  {Thomas}(2019)}]{Jaiswal_2019}%
  \BibitemOpen
  \bibfield  {author} {\bibinfo {author} {\bibfnamefont {S.}~\bibnamefont
  {Jaiswal}}\ and\ \bibinfo {author} {\bibfnamefont {E.}~\bibnamefont
  {Thomas}},\ }\href {\doibase 10.1088/2516-1067/ab1f30} {\bibfield  {journal}
  {\bibinfo  {journal} {Plasma Research Express}\ }\textbf {\bibinfo {volume}
  {1}},\ \bibinfo {pages} {025014} (\bibinfo {year} {2019})}\BibitemShut
  {NoStop}%
\bibitem [{\citenamefont {Wang}\ \emph {et~al.}(2019)\citenamefont {Wang},
  \citenamefont {Huang},\ and\ \citenamefont {Feng}}]{Wang19}%
  \BibitemOpen
  \bibfield  {author} {\bibinfo {author} {\bibfnamefont {K.}~\bibnamefont
  {Wang}}, \bibinfo {author} {\bibfnamefont {D.}~\bibnamefont {Huang}}, \ and\
  \bibinfo {author} {\bibfnamefont {Y.}~\bibnamefont {Feng}},\ }\href {\doibase
  10.1103/PhysRevE.99.063206} {\bibfield  {journal} {\bibinfo  {journal} {Phys.
  Rev. E}\ }\textbf {\bibinfo {volume} {99}},\ \bibinfo {pages} {063206}
  (\bibinfo {year} {2019})}\BibitemShut {NoStop}%
\bibitem [{\citenamefont {Himpel}\ and\ \citenamefont
  {Melzer}(2019)}]{Himpel19}%
  \BibitemOpen
  \bibfield  {author} {\bibinfo {author} {\bibfnamefont {M.}~\bibnamefont
  {Himpel}}\ and\ \bibinfo {author} {\bibfnamefont {A.}~\bibnamefont
  {Melzer}},\ }\href {\doibase 10.1103/PhysRevE.99.063203} {\bibfield
  {journal} {\bibinfo  {journal} {Phys. Rev. E}\ }\textbf {\bibinfo {volume}
  {99}},\ \bibinfo {pages} {063203} (\bibinfo {year} {2019})}\BibitemShut
  {NoStop}%
\bibitem [{\citenamefont {Choi}\ \emph {et~al.}(2019)\citenamefont {Choi},
  \citenamefont {Dharuman},\ and\ \citenamefont {Murillo}}]{Choi19}%
  \BibitemOpen
  \bibfield  {author} {\bibinfo {author} {\bibfnamefont {Y.}~\bibnamefont
  {Choi}}, \bibinfo {author} {\bibfnamefont {G.}~\bibnamefont {Dharuman}}, \
  and\ \bibinfo {author} {\bibfnamefont {M.~S.}\ \bibnamefont {Murillo}},\
  }\href {\doibase 10.1103/PhysRevE.100.013206} {\bibfield  {journal} {\bibinfo
   {journal} {Phys. Rev. E}\ }\textbf {\bibinfo {volume} {100}},\ \bibinfo
  {pages} {013206} (\bibinfo {year} {2019})}\BibitemShut {NoStop}%
\bibitem [{\citenamefont {Lisina}\ \emph {et~al.}(2019)\citenamefont {Lisina},
  \citenamefont {Vaulina},\ and\ \citenamefont {Lisin}}]{Lisina19}%
  \BibitemOpen
  \bibfield  {author} {\bibinfo {author} {\bibfnamefont {I.~I.}\ \bibnamefont
  {Lisina}}, \bibinfo {author} {\bibfnamefont {O.~S.}\ \bibnamefont {Vaulina}},
  \ and\ \bibinfo {author} {\bibfnamefont {E.~A.}\ \bibnamefont {Lisin}},\
  }\href {\doibase 10.1103/PhysRevE.99.013207} {\bibfield  {journal} {\bibinfo
  {journal} {Phys. Rev. E}\ }\textbf {\bibinfo {volume} {99}},\ \bibinfo
  {pages} {013207} (\bibinfo {year} {2019})}\BibitemShut {NoStop}%
\bibitem [{\citenamefont {Sukhinin}\ \emph {et~al.}(2017)\citenamefont
  {Sukhinin}, \citenamefont {Fedoseev}, \citenamefont {Salnikov}, \citenamefont
  {Rostom}, \citenamefont {Vasiliev},\ and\ \citenamefont
  {Petrov}}]{Sukhinin17}%
  \BibitemOpen
  \bibfield  {author} {\bibinfo {author} {\bibfnamefont {G.~I.}\ \bibnamefont
  {Sukhinin}}, \bibinfo {author} {\bibfnamefont {A.~V.}\ \bibnamefont
  {Fedoseev}}, \bibinfo {author} {\bibfnamefont {M.~V.}\ \bibnamefont
  {Salnikov}}, \bibinfo {author} {\bibfnamefont {A.}~\bibnamefont {Rostom}},
  \bibinfo {author} {\bibfnamefont {M.~M.}\ \bibnamefont {Vasiliev}}, \ and\
  \bibinfo {author} {\bibfnamefont {O.~F.}\ \bibnamefont {Petrov}},\ }\href
  {\doibase 10.1103/PhysRevE.95.063207} {\bibfield  {journal} {\bibinfo
  {journal} {Phys. Rev. E}\ }\textbf {\bibinfo {volume} {95}},\ \bibinfo
  {pages} {063207} (\bibinfo {year} {2017})}\BibitemShut {NoStop}%
\bibitem [{\citenamefont {Hartmann}\ \emph {et~al.}(2017)\citenamefont
  {Hartmann}, \citenamefont {Reyes}, \citenamefont {Korolov}, \citenamefont
  {Matthews},\ and\ \citenamefont {Hyde}}]{Hartmann17}%
  \BibitemOpen
  \bibfield  {author} {\bibinfo {author} {\bibfnamefont {P.}~\bibnamefont
  {Hartmann}}, \bibinfo {author} {\bibfnamefont {J.~C.}\ \bibnamefont {Reyes}},
  \bibinfo {author} {\bibfnamefont {I.}~\bibnamefont {Korolov}}, \bibinfo
  {author} {\bibfnamefont {L.~S.}\ \bibnamefont {Matthews}}, \ and\ \bibinfo
  {author} {\bibfnamefont {T.~W.}\ \bibnamefont {Hyde}},\ }\href {\doibase
  10.1063/1.4985080} {\bibfield  {journal} {\bibinfo  {journal} {Physics of
  Plasmas}\ }\textbf {\bibinfo {volume} {24}},\ \bibinfo {pages} {060701}
  (\bibinfo {year} {2017})}\BibitemShut {NoStop}%
\bibitem [{\citenamefont {Spreiter}\ and\ \citenamefont
  {Walter}(1999)}]{SPREITER1999102}%
  \BibitemOpen
  \bibfield  {author} {\bibinfo {author} {\bibfnamefont {Q.}~\bibnamefont
  {Spreiter}}\ and\ \bibinfo {author} {\bibfnamefont {M.}~\bibnamefont
  {Walter}},\ }\href {\doibase https://doi.org/10.1006/jcph.1999.6237}
  {\bibfield  {journal} {\bibinfo  {journal} {Journal of Computational
  Physics}\ }\textbf {\bibinfo {volume} {152}},\ \bibinfo {pages} {102 }
  (\bibinfo {year} {1999})}\BibitemShut {NoStop}%
\bibitem [{\citenamefont {Semenov}\ \emph {et~al.}(2015)\citenamefont
  {Semenov}, \citenamefont {Khrapak},\ and\ \citenamefont
  {Thomas}}]{Khrapak15}%
  \BibitemOpen
  \bibfield  {author} {\bibinfo {author} {\bibfnamefont {I.~L.}\ \bibnamefont
  {Semenov}}, \bibinfo {author} {\bibfnamefont {S.~A.}\ \bibnamefont
  {Khrapak}}, \ and\ \bibinfo {author} {\bibfnamefont {H.~M.}\ \bibnamefont
  {Thomas}},\ }\href {\doibase 10.1063/1.4935846} {\bibfield  {journal}
  {\bibinfo  {journal} {Physics of Plasmas}\ }\textbf {\bibinfo {volume}
  {22}},\ \bibinfo {pages} {114504} (\bibinfo {year} {2015})}\BibitemShut
  {NoStop}%
\bibitem [{\citenamefont {Ott}\ \emph {et~al.}(2014)\citenamefont {Ott},
  \citenamefont {L\"owen},\ and\ \citenamefont {Bonitz}}]{Ott14}%
  \BibitemOpen
  \bibfield  {author} {\bibinfo {author} {\bibfnamefont {T.}~\bibnamefont
  {Ott}}, \bibinfo {author} {\bibfnamefont {H.}~\bibnamefont {L\"owen}}, \ and\
  \bibinfo {author} {\bibfnamefont {M.}~\bibnamefont {Bonitz}},\ }\href
  {\doibase 10.1103/PhysRevE.89.013105} {\bibfield  {journal} {\bibinfo
  {journal} {Phys. Rev. E}\ }\textbf {\bibinfo {volume} {89}},\ \bibinfo
  {pages} {013105} (\bibinfo {year} {2014})}\BibitemShut {NoStop}%
\bibitem [{\citenamefont {Ott}\ \emph {et~al.}(2015)\citenamefont {Ott},
  \citenamefont {Bonitz},\ and\ \citenamefont {Donk\'o}}]{Ott15}%
  \BibitemOpen
  \bibfield  {author} {\bibinfo {author} {\bibfnamefont {T.}~\bibnamefont
  {Ott}}, \bibinfo {author} {\bibfnamefont {M.}~\bibnamefont {Bonitz}}, \ and\
  \bibinfo {author} {\bibfnamefont {Z.}~\bibnamefont {Donk\'o}},\ }\href
  {\doibase 10.1103/PhysRevE.92.063105} {\bibfield  {journal} {\bibinfo
  {journal} {Phys. Rev. E}\ }\textbf {\bibinfo {volume} {92}},\ \bibinfo
  {pages} {063105} (\bibinfo {year} {2015})}\BibitemShut {NoStop}%
\bibitem [{\citenamefont {Begum}\ and\ \citenamefont {Das}(2016)}]{Begum16}%
  \BibitemOpen
  \bibfield  {author} {\bibinfo {author} {\bibfnamefont {M.}~\bibnamefont
  {Begum}}\ and\ \bibinfo {author} {\bibfnamefont {N.}~\bibnamefont {Das}},\
  }\href {\doibase 10.9790/4861-0806024954} {\bibfield  {journal} {\bibinfo
  {journal} {IOSR Journal of Applied Physics}\ }\textbf {\bibinfo {volume}
  {8}},\ \bibinfo {pages} {49} (\bibinfo {year} {2016})}\BibitemShut {NoStop}%
\bibitem [{\citenamefont {Karasev}\ \emph {et~al.}(2019)\citenamefont
  {Karasev}, \citenamefont {Dzlieva}, \citenamefont {D'yachkov}, \citenamefont
  {Novikov}, \citenamefont {Pavlov},\ and\ \citenamefont
  {Tarasov}}]{Karasev19}%
  \BibitemOpen
  \bibfield  {author} {\bibinfo {author} {\bibfnamefont {V.}~\bibnamefont
  {Karasev}}, \bibinfo {author} {\bibfnamefont {E.}~\bibnamefont {Dzlieva}},
  \bibinfo {author} {\bibfnamefont {L.}~\bibnamefont {D'yachkov}}, \bibinfo
  {author} {\bibfnamefont {L.}~\bibnamefont {Novikov}}, \bibinfo {author}
  {\bibfnamefont {S.}~\bibnamefont {Pavlov}}, \ and\ \bibinfo {author}
  {\bibfnamefont {S.}~\bibnamefont {Tarasov}},\ }\href {\doibase
  10.1002/ctpp.201800136} {\bibfield  {journal} {\bibinfo  {journal}
  {Contributions to Plasma Physics}\ }\textbf {\bibinfo {volume} {59}},\
  \bibinfo {pages} {e201800136} (\bibinfo {year} {2019})}\BibitemShut {NoStop}%
\end{thebibliography}%

\end{document}